\documentclass{article}
\usepackage{amssymb}
\usepackage{fleqn}
\usepackage{epsfig}
\usepackage{graphicx}
\usepackage{amsmath}
\def\beq{\begin{equation}}
\def\eeq{\end{equation}}
\def\bea{\begin{eqnarray}}
\def\eea{\end{eqnarray}}
\def\bq{\begin{quote}}
\def\eq{\end{quote}}

\def\gappeq{\mathrel{\rlap {\raise.5ex\hbox{$>$}}
{\lower.5ex\hbox{$\sim$}}}}
\def\lappeq{\mathrel{\rlap{\raise.5ex\hbox{$<$}}
{\lower.5ex\hbox{$\sim$}}}}
\def\Toprel#1\over#2{\mathrel{\mathop{#2}\limits^{#1}}}

\begin{document}

\pagestyle{empty}
\begin{flushright}
ROME1/1407/05~

DSFNA1/25/2005
\end{flushright}
\vspace*{15mm}

\begin{center}
\textbf{
THRESHOLD RESUMMED SPECTRA IN $B\rightarrow X_u l \nu$ DECAYS IN NLO (I)
} \\[0pt]

\vspace*{1cm}

\textbf{Ugo Aglietti}\footnote{e-mail address: Ugo.Aglietti@roma1.infn.it} \\[0pt]

\vspace{0.3cm}
Dipartimento di Fisica,\\
Universit\`a di Roma ``La Sapienza'', \\
and I.N.F.N.,
Sezione di Roma, Italy. \\[1pt]

\vspace{0.3cm}
\textbf{Giulia Ricciardi}\footnote{e-mail address:
Giulia.Ricciardi@na.infn.it} \\ [0pt]

\vspace{0.3cm} Dipartimento di Scienze Fisiche,\\
Universit\`a di Napoli ``Federico II'' \\
and I.N.F.N.,
Sezione di Napoli, Italy. \\ [1pt]

\vspace{0.3cm}
\textbf{Giancarlo Ferrera}\footnote{e-mail address: Giancarlo.Ferrera@roma1.infn.it} \\[0pt]

\vspace{0.3cm}
Dipartimento di Fisica,\\
Universit\`a di Roma ``La Sapienza'', \\
and I.N.F.N.,
Sezione di Roma, Italy. \\[1pt]

%$~~~$ \\[0pt]
\vspace*{1cm} \textbf{Abstract} \\[0pt]
\end{center}
We evaluate threshold resummed spectra in $B\rightarrow X_u l \nu$
decays in next-to-leading order. We present results for the
distribution in the hadronic variables $E_X$ and $m_X^2/E_X^2$,
for the distribution in $E_X$ and for the distribution in $E_X$
and $E_l$, where $E_X$ and $m_X$ are the total energy and the
invariant mass of the final hadronic state $X_u$ respectively and
$E_l$ is the energy of the charged lepton. We explicitly show that
all these spectra (where there is no integration over the hadronic
energy) can be directly related to the photon spectrum in
$B\rightarrow X_s \gamma$ via short-distance coefficient
functions.

\vspace*{4cm} \noindent %\rule[.1in]{16.5cm}{.002in}

\vfill\eject
%\pagestyle{empty}
%\clearpage\mbox{}\clearpage

\setcounter{page}{1} \pagestyle{plain}

\section{Introduction and summary of the results}

A long-standing problem in particle physics is the understanding
of strong interactions at low energies. While at very low energies,
of the order of the hadronic scale $\Lambda\approx 300$ MeV,
perturbative QCD is of no use and alternative methods
have been developed in decades (such as
quark models, chiral lagrangians, lattice QCD, etc.),
at intermediate energies, of the order of a few GeV,
perturbative computations can be combined with non-perturbative
models to predict a variety of cross sections and decay rates.
Among these moderate hard scale phenomena is beauty physics,
which is indeed characterized by a hard scale of a few GeV.
The measured decay spectra often receive large contributions
at the endpoints ---
in the case of the hadron energy spectrum, in the middle of the domain ---
from long-distance effects related to soft interactions between
the heavy quark and the light degrees of freedom.

The main non perturbative effect is the well-known Fermi motion, which
classically can be described as a small vibration of the heavy quark
inside the $B$ meson because of the momentum exchange with the valence
quark; in the quantum theory it is also the virtuality of the heavy
quark that matters. This effect is important in the end-point region,
because it produces some smearing of the partonic spectra.

These long distance effects manifest themselves in perturbation theory
in the form of series of large infrared logarithms,
coming from an ``incomplete'' cancellation of infrared
divergencies in real and virtual diagrams.
The probability for instance for a light quark produced
in a process with a hard scale $Q$ to evolve into a jet with an
invariant mass smaller than $m$ is written in leading order as \cite{cattren1}:
\begin{eqnarray}
\label{tobeginwith}
J(m) & = & 1\, + \, A_1 \, \alpha_S \, \int_0^1 \frac{d\omega}{\omega}
\int_0^1 \frac{d\theta^2}{\theta^2} \, \Theta\left(\frac{m^2}{Q^2}\,-\,\omega\,\theta^2\right)
\, - \, \, A_1 \, \alpha_S \, \int_0^1 \frac{d\omega}{\omega} \int_0^1 \frac{d\theta^2}{\theta^2}
\nonumber\\
&=& 1 \, - \, \frac{A_1}{2}\,\alpha_S \, \log^2\left(\frac{Q^2}{m^2}\right),
\end{eqnarray}
where $\omega$ is the energy of a gluon emitted by the light quark normalized to the
hard scale, $\theta$ is its emission angle and $A_1$ is a positive constant
(see sec.~\ref{secthresum}).
The first integral on the r.h.s. is the real contribution while the second integral
is the virtual one.
Both integrals are separately divergent for $\omega=0$
 --- soft singularity --- as well as for $\theta=0$ --- collinear singularity,
but their sum is finite.
``Complete'' real-virtual cancellation occurs only for $m=Q$, i.e. in the
completely inclusive evolution of the quark line, while for $m<Q$
there is a left-over double logarithm because of the smaller integration
region of the real diagrams.
Multiple gluon emission occurs in higher orders of perturbation theory;
it can be described as a classical branching process and gives rise to the
double logarithmic series,
i.e. to powers of the last term $\alpha_S\log^2\left(Q^2/m^2\right)$
on the r.h.s. of eq.~(\ref{tobeginwith}) \cite{cattren1,dok}.

We may say that perturbation theory ``signals'' long-distance
effects in a specific way --- even though a quantitative description
of the latter has to include also some truly non-perturbative
component.
A theoretical study of the universality of these long-distance
effects can therefore be done {\it inside} perturbation theory,
by comparing the logarithmic structure of different distributions.
In other words, if these long-distance effects are universal,
this has certainly to show up in perturbation theory: things
have to work in perturbation theory first.
The aim of this work is to study the relation of long-distance effects
between different distributions by means of resummed perturbation
theory.

In general, let us consider the semi-inclusive decays
\begin{equation}
\label{seminclusive}
B \, \rightarrow \, X_q \, + \, {\rm (non~QCD~partons)},
\end{equation}
where $X_q$ is any hadronic final state coming from the fragmentation
of the light quark $q=u,d,s$ and the non QCD partons are typically
a photon, a lepton-neutrino pair, a lepton-antilepton pair, etc.
This system of particle(s), with total four-momentum $q_{\mu}$,
constitutes a ``probe'' for the hadronic process,
as in the case of deep-inelastic-scattering (DIS) of leptons off hadrons.
Without any generality loss, we can work in the $b$ rest frame,
where $p_b^{\mu}=m_b v^{\mu}$, with $m_b$ being the beauty mass
and $v^{\mu}=(1;0,0,0)$ being the classical 4-velocity.
The hadronic subprocess in (\ref{seminclusive}) is characterized
by the following three scales:
\begin{equation}
m_b,~~~~~E_X~~~~{\rm and}~~~~m_X~~~~~~~~~~~~~~~~~~~~(m_b\ge E_X),
\end{equation}
where $m_X$ and $E_X$ are the invariant mass and the total energy
of the final hadronic state $X_q$, respectively.
We are interested in the so-called threshold region, which can be
defined in all generality as the one having
\begin{equation}
\label{inhibited}
m_X \, \ll \, E_X.
\end{equation}
The region (\ref{inhibited}) is sometimes called radiation-inhibited,
because the emitted radiation naturally produces final states with
an invariant mass of the order of the hard scale: $m_X \sim O(E_X)$.
It is also called semi-inclusive because experimentally,
to satisfy the constraint (\ref{inhibited}), most hadronic final states
have to be discarded.

The processes we are going to consider are the well-known radiative
decay with a real photon in the final state,
\begin{equation}
\label{bsgamma}
B \, \rightarrow \, X_s \, + \, \gamma
\end{equation}
and the semi-leptonic decay,
\footnote{
The results for the semileptonic decay are easily extended to the
radiative decay with the photon converting into a lepton
pair,
\begin{equation}
\label{veryrare}
B \, \rightarrow \, X_s \, + \, l \, + \, \overline{l}.
\end{equation}
}

\begin{equation}
\label{semilep}
B \, \rightarrow \, X_u \, + \, l \, + \, \nu.
\end{equation}

In perturbative QCD, the hadronic subprocess in (\ref{seminclusive})
consists of a heavy quark
decaying into a light quark which evolves later
into a jet of soft and collinear
partons because of infrared divergencies.
In leading order, one only considers the emission of soft
gluons at small angle by the light quark (see eq.~(\ref{tobeginwith}));
the final state $X_q$
consists of a jet with the leading (i.e. most energetic) quark $q$
originating the jet itself.
In next-to-leading order one has to take into account two
different single-logarithmic effects:
$(a)$ hard emission at small angle by the light quark $q$
and $(b)$ soft emission at large angle by the heavy quark.
Because of $(a)$, the final state consists of a jet with many hard
partons and, in general, the leading parton is no longer the quark $q$
which originated the jet itself.
Because of $(b)$, the final state does not contain only an isolated jet,
but also soft partons in any space direction.
The main result of \cite{me} is that the large threshold logarithms appearing
in (\ref{seminclusive}) are conveniently organized as a series of the
form:
\begin{eqnarray}
\label{finallargelogs}
&&
\sum_{n=1}^{\infty}
\sum_{k=1}^{2n}
c_{n k} \,\alpha^n(Q) \, \log^k \frac{Q^2}{m_X^2}
\nonumber\\
&=& c_{12}\,\alpha(Q)\,\log^2\frac{Q^2}{m_X^2}
\, + \, c_{11} \, \alpha(Q)\,\log\frac{Q^2}{m_X^2}
\, + \, c_{24} \, \alpha^2(Q)\,\log^4\frac{Q^2}{m_X^2}
\, + \, c_{23} \, \alpha^2(Q)\,\log^3\frac{Q^2}{m_X^2} \, + \, \cdots\cdots,
\end{eqnarray}
where $\alpha(Q) = \alpha_S(Q)$ is the QCD coupling
and the hard scale $Q$ is determined by the final hadronic energy $E_X$
\footnote{
The factor two is inserted in such a way that the hard
scale coincides with $m_b$  in the radiative decay (see later).
The essential point however is that $Q$ is proportional
to $E_X$ via a proportionality constant of order one, whose precise value
is irrelevant.}:
\begin{equation}
\label{basicrel}
Q \, = \, 2 E_X.
\end{equation}
These large logarithms are factorized into a universal QCD form factor.
Let us summarize the derivation of (\ref{finallargelogs}) and (\ref{basicrel}).
We take the infinite mass limit for the beauty quark
while keeping the hadronic energy and the hadronic mass fixed
\footnote{This limit has not to be confused with that one relevant
for the shape function, also called structure function
of the heavy flavors, which is $E_X\rightarrow\infty$, $m_X\rightarrow\infty$
with $m_X^2/E_X\rightarrow {\rm const}$ (the latter implies
$m_b\rightarrow\infty$, but the converse is not true).
The shape function describes soft interactions only and therefore does not factorize
the whole logarithmic structure, missing the large logarithms coming
from hard collinear emission off the light quark
%\cite{coefun}
.}:
\begin{equation}
\label{solomb}
m_b \, \rightarrow \, \infty,~~~~~{\rm with}~~~~~
E_X \,\, \, {\rm and} \,\, \, m_X\,\rightarrow\,{\rm const}.
\end{equation}
This takes us into an effective theory in which the
beauty quark is replaced by a static quark, as recoil effects
are neglected in the limit (\ref{solomb}).
If we write the beauty quark momentum as $p_b=m_b v+k$, where
$k$ is a soft
momentum, the infinite mass limit of the propagator is easily
obtained as:
\begin{equation}
\label{statlim}
S_F(p)\, = \, \left(\frac{1+\hat{v}}{2}\,+\,\frac{\hat{k}}{2m}\right)
\frac{1}{v\cdot k \,+ \, k^2/(2m) \, + \, i \epsilon}
\,\,\,\,\, \rightarrow \,\,\,\,\,
\frac{1\, + \, \hat{v}}{2}\,\frac{1}{v\cdot k \, + \, i \epsilon}
~~~~~~~~{\rm (static~limit),}
\end{equation}
where $\hat{a}\equiv\gamma_{\mu}a^{\mu}$.
As discussed above, the beauty quark contributes to the QCD form
factor via large logarithms coming from soft emissions,
which are correctly described by a static quark.
Since the light quark propagator is not touched by the limit
(\ref{solomb}), we conclude that all soft and/or collinear
emissions are correctly described by this limit.
Since the heavy flavor mass has disappeared with the limit (\ref{solomb}),
the only remaining scales in the hadronic subprocess are $m_X$ and $E_X$.
Only one adimensional quantity can be constructed out of them, for example
the ratio $E_X/m_X$, which is therefore the only possible argument for the large
logarithms, in agreement with (\ref{finallargelogs}).
Furthermore, the hard scale $Q$ is given by the greatest scale in the game, i.e.
by the hadronic energy $E_X$, in agreement with (\ref{basicrel}).

The argument given above, however, is not rigorous: let us refine it.
The limit (\ref{solomb}) is indeed singular in quantum field theory:
one cannot remove degrees of freedom without paying some price.
Let us consider for simplicity's sake the semileptonic decay
(\ref{semilep}), even though the conclusions are general.
The vector and axial-vector currents
responsible for the $b\rightarrow u$ transition are conserved or
partially conserved in QCD, implying that the $O(\alpha)$ virtual
corrections are ultraviolet finite.
These corrections contain however terms of the form
\begin{equation}
\label{logmb}
\gamma_0 \, \alpha\,\log \frac{m_b}{E_X}~~~~~~~~~~~~~~~~{\rm (ordinary~QCD)},
\end{equation}
which diverge in the limit (\ref{solomb}) ($\gamma_0$ is a constant).
If one takes the limit (\ref{solomb}) {\it ab initio}, i.e. before integrating
the loop, some divergence
is expected in the loop integrals, as it is indeed the case.
Technically, that occurs because
the static propagator is of the form $1/(k_0+i\epsilon)$ (see eq.~(\ref{statlim})) and,
unlike the ordinary propagator, has no damping for $|\vec{k}|\rightarrow\infty$.
It can be shown that the $b\rightarrow u$ vector and axial-vector currents
are no more conserved or partially conserved in the static theory.
Therefore, unlike the QCD case, the $O(\alpha)$ virtual corrections are
ultraviolet divergent in the static theory and produce,
after renormalization, terms corresponding to (\ref{logmb})
of the form
\begin{equation}
\gamma_0 \, \alpha\,\log \frac{\mu}{E_X}~~~~~~~~~~~~~~~~{\rm (effective~theory)},
\end{equation}
in which basically the heavy flavor mass $m_b$ is replaced
by the renormalization point $\mu$ --- the coefficient $\gamma_0$
being the same.
The hadronic subprocess in the static theory therefore has
amplitudes depending on the physical scales $E_X$ and $m_X$ as well as
on the renormalization scale $\mu$.
If we neglect terms suppressed by inverse powers of the beauty mass
$\sim 1/m_b^n$, we have that the physical scale $m_b$ is replaced by the
renormalization point $\mu$ in the effective theory: $m_b \rightarrow \mu$.
The effective currents $\tilde{J}_{\nu}$ and the coupling constant $\alpha$
are renormalized at the scale $\mu$:
$\tilde{J}_{\nu}=\tilde{J}_{\nu}(\mu)$ and $\alpha=\alpha(\mu)$.
The effective amplitudes contain terms of the form $\alpha^n\log^k \mu/E_X$ $(k\le n)$,
which are large logarithms for $\mu\gg E_X$ or $\mu\ll E_X$.
To have convergence of the perturbative series, the large logarithms
above must be resummed by taking $\mu=O(E_X)$, i.e.
$\mu = k \, E_X$ with $k=O(1)$.
This implies that the effective currents and the coupling are
evaluated at a scale of the order of the hadronic energy:
$\tilde{J}_{\nu}=\tilde{J}_{\nu}(k\,E_X)$ and $\alpha=\alpha(k\,E_X)$.
We have therefore proved that the hard scale $Q$ is fixed by the
final hadronic energy $E_X$ and not by the beauty mass $m_b$:
\begin{equation}
Q \, = \, \mu\, = \, k \, E_X~~~~~~~~~~{\rm with}~k \, = \, O(1).
\end{equation}

\noindent
Let us go back to the general process (\ref{seminclusive}).
Kinematics gives:
\begin{equation}
2 E_X \, = \, m_b
\left(
1 \, - \, \frac{q^2}{m_b^2} \, + \, \frac{m_X^2}{m_b^2}
\right).
\end{equation}
The simplest processes are those with a light-like probe, i.e. with $q^2=0$,
where
\begin{equation}
2 E_X \, = \, m_b
\left(
1 \, + \, \frac{m_X^2}{m_b^2}
\right) \, \simeq \, m_b.
\end{equation}
This case corresponds to the radiative decay (\ref{bsgamma}).
In this case, the final hadronic energy is always large and of the order
of the heavy-flavor mass:
\begin{equation}
Q \, \approx \, m_b~~~~~~~~{\rm (radiative~decay)}.
\end{equation}
On the other hand, in the semi-leptonic decay (\ref{semilep})
\footnote{
The same is also true for the radiative decay with the photon converting
into a lepton pair (\ref{veryrare}).},
the lepton pair can have a large
invariant mass,
\begin{equation}
q^2 \, \sim \, O\left(m_b^2\right),
\end{equation}
implying a substantial reduction of the hard scale:
\begin{equation}
Q \, \ll \, m_b.
\end{equation}
This fact is one of the complications in the threshold resummation
of the semileptonic decay spectra:
while in the radiative decay (\ref{bsgamma}), the hard scale $Q$ is
always large in the threshold region, and of the order of
$m_b$, this is no longer true in the semileptonic decay.
The hadronic subprocesses have in general different hard
scales in the two decays. If one integrates over $q^2$,
for example because of undetected neutrino momentum, there is a mixing
of hadronic contributions with different hard scales in the semileptonic
case.
However, it turns out by explicit computation
that the contributions from a large $q^2$, i.e.
with a small hard scale in the hadronic subprocess,
are rather suppressed (see sec.~\ref{hadrvars}).

At fixed $Q$, the large logarithms in (\ref{seminclusive})
can be factorized into a QCD form factor, which
is universal in the sense that it depends only on the hadronic subprocess.
The differences between, let us say, the radiative decay (\ref{bsgamma})
and the semileptonic decay (\ref{semilep}) only enter in the specific
form of a short-distance coefficient function multiplying the QCD
form factor (and in the form of a remainder function collecting
non factorized, small contributions, see next section).

The discussion above can be summarized as follows.
The hard scale $Q = 2\,E_X$ in (\ref{seminclusive})
appears in the argument in the infrared
logarithms as well as in the argument of the running coupling.
In the radiative decay, because of kinematics, the hard scale
is always large and of the order of the beauty mass: $Q\approx m_b$, while
in the semileptonic case kinematical configurations are possible with
$Q\approx m_b$ as well as with $Q \ll m_b$.
The main complication in semileptonic decays is that by performing
kinematical integrations (for example over the neutrino energy), one
may integrate over the hard scale of the hadronic subprocess.
While in radiative decays the hard scale is fixed, in the semileptonic
decays there can be a mixing of different hadronic subprocesses.
A non-trivial picture of some semileptonic
decay spectra emerges: there are long-distance effects
which cannot be extracted by the radiative decay, related
to a small final hadronic energy, but their effect turns out to be
small at the end because of a kinematical suppression
of the states with a small hard scale.
The decay spectra in (\ref{semilep})
can therefore be divided into two classes:
\begin{enumerate}
\item
distributions in which the hadronic energy $E_X$ is not integrated over.
These distributions can be related via short-distance
coefficients to the photon spectrum in the radiative decay
(\ref{bsgamma}).
In particular, the structure of the threshold logarithms is the
same as in decay (\ref{bsgamma}).
In this paper we restrict ourselves to these simpler distributions;
\item
distributions in which the hadronic energy is integrated over and
therefore all the hadronic energies contribute.
These are for instance the hadron mass distribution or the charged
lepton energy distribution. In all these cases,
the structure of the threshold logarithms is different
from that one in (\ref{bsgamma}) and by far more complicated.
The analysis of some of these distributions, which present novel features with
respect to $B\rightarrow X_s\gamma$, is given in \cite{wip}.
\end{enumerate}

Let us make a simple analogy with $e^+e^-$ annihilation
into hadrons. In the center-of-mass (c.o.m.) frame, the final state consists
of a $q\overline{q}$ pair, which are emitted back to back
with a high virtuality and evolve later into two jets:
\begin{equation}
\label{epluseminus}
e^+ \, + \, e^- \, \rightarrow \, q \, + \, \overline{q}
\, \rightarrow \, J_q \, + \, J_{\overline{q}}.
\end{equation}
Roughly speaking, the final state $X_q$ in (\ref{seminclusive}), consisting
in a single jet, is ``half'' of that in (\ref{epluseminus}), consisting
of the two jets $J_q$ and  $J_{\overline{q}}$.
Deviations from this independent fragmentation picture arise in next-to-leading order because
of large-angle soft emission by the heavy quark in (\ref{seminclusive}),
which has no analogue in (\ref{epluseminus}).
The structure of $e^+e^-$ hadronic final states is
conveniently analyzed by means of so-called shape variables,
one of the most studied being the heavy jet mass $m_H^2$, defined as
\begin{equation}
m_H^2 \, = \, \max\left\{ m_R^2, m_L^2 \right\},
\end{equation}
where $m_R$ and $m_L$ are the invariant masses of the particles
in the right and left hemispheres of the event respectively.
The hemispheres are defined cutting the space with a plane
orthogonal to the thrust axis $\vec{n}$, the latter
defined as the direction maximizing
\begin{equation}
\sum_i |\vec{p}_i \cdot \hat{n}|,
\end{equation}
i.e. basically the sum of length of longitudinal momenta.
The sum extends over all hadrons --- partons in the perturbative
computation.
For $m_H \ll Q$, where $Q$ is the hard scale to be identified here with
the c.o.m. energy, hard emission at large angle by the
$q\overline{q}$ pair cannot occur and the final state consists
of two narrow jets around the original $q\overline{q}$ direction,
which can be identified with $\vec{n}$.
The $O(\alpha)$ computation gives large logarithms
of similar form to those in (\ref{finallargelogs}) \cite{thrust}:
\begin{equation}
\alpha(Q)\,\log^2 \frac{Q^2}{m_H^2}~~~~{\rm and}~~~~\alpha(Q)\,\log \frac{Q^2}{m_H^2}.
\end{equation}
There is not a simple relation between, let us say, the heavy jet mass
distribution at the $Z^0$ peak,
\begin{equation}
\label{former}
\frac{d\sigma}{dm_H^2}\left(Q \, = \, m_Z\right)
\end{equation}
and the integral of this quantity over
$Q$ from a small energy $\epsilon\sim m_H$
up to $m_Z$ with some weight function $\phi(Q)$:
\begin{equation}
\label{latter}
\frac{d\hat{\sigma}}{dm_H^2} \, = \,
\int_{\epsilon}^{m_Z} \, d Q \, \phi(Q) \, \frac{d\sigma}{dm_H^2}
\left( Q \right).
\end{equation}
Radiative $B$ decays (\ref{bsgamma}) and semileptonic spectra
(\ref{semilep}) in class 1. are the analog of the former
distribution (\ref{former}), while semileptonic spectra in class
2. are the analog of the latter case (\ref{latter}). The analog of
the suppression in the semileptonic spectra 2. of the
contributions from large $q^2$ is the suppression of the weight
function $\phi(Q)$ for $Q\,\ll\, m_Z$.

Many properties of the distributions we are going to derive in
this work can be understood with a qualitative discussion on the hadron
energy spectrum,
\begin{equation}
\frac{d \, \Gamma}{d\,E_X},
\end{equation}
which exhibits a remarkable phenomenon related to the occurrence
of infrared singularities inside the physical domain, instead than
at the boundary as it is usually the case.
This phenomenon has been studied in the framework of jet physics
and is known as the ``Sudakov shoulder'' \cite{sudshould,enspec,ndf,me}.
Let us discuss it in the present case in physical terms.
In lowest order, the semileptonic decay (\ref{semilep}) involves three massless partons
in the final state:
\begin{equation}
b \, \rightarrow \, u \, + \, l \, + \, \nu.
\end{equation}
According to kinematics, any final state parton can take at most
half of the initial energy, implying that
\begin{equation}
\label{kinlimit}
E_X^{(0)} \, = \, E_u \, \le \, \frac{m_b}{2}.
\end{equation}
In lowest order, the final hadronic state consists indeed of the $up$ quark only:
$X_u=u$.
To order $\alpha$, a real gluon is radiated and the final hadronic state is
a two-particle system: $X_u=u+g$.
The final hadronic energy is not restricted anymore to half the beauty mass
but can go up to the whole beauty mass:
\begin{equation}
E_X^{(1)} \, = \, E_u \, + \, E_g \, \le \, m_b.
\end{equation}
For example, just consider an energetic {\it up} quark recoiling
against the gluon, with a soft electron and a soft neutrino.
The relevant case for us is a final state with the
$up$ quark of energy $\approx m_b/2$ and a soft and/or a collinear gluon.
Such a state has a total energy slightly above $m_b/2$ and the
matrix element is logarithmically enhanced because of the
well-known infrared singularities.
Such logarithmic enhancement cannot be cancelled by the $O(\alpha)$
virtual corrections, because of their tree-level
kinematical limitation (\ref{kinlimit}).
We are left therefore with large infrared logarithms, of the
form
\begin{equation}
\label{logsintro}
~~~~~~~~~~\alpha\,\log^2\left(E_X-\frac{m_b}{2}\right),~~~~~\alpha \, \log\left(E_X - \frac{m_b}{2}\right)
~~~~~~~~~~~~~~~~~~~\left(E_X\ge\frac{m_b}{2}\right),
\end{equation}
which are final and produce an infrared divergence for $E_X\rightarrow +\,m_b/2$.
On the other hand, for $E_X < m_b/2$ there are no large logarithms
of the form $\alpha\log^k(m_b/2-E_X)$ $(k=1,2)$, because in this case real-virtual
cancellation may occur, and it actually does.
Let us summarize:
the $O(\alpha)$ spectrum has an infrared singularity right
in the middle of the domain, for $\overline{E}_X=m_b/2$, because the
lowest order spectrum has a discontinuity in this point, above
which it vanishes identically because of kinematics.

\noindent
This infrared singularity is integrable, as
\begin{equation}
\int_{m_b/2}^{m_b/2+\delta} d E_X \, \alpha \, \log^k\left(E_X-\frac{m_b}{2}\right)
\, < \, \infty,
\end{equation}
where $\delta>0$ is some energy-resolution parameter.
The infrared divergence is therefore eliminated with some smearing over the hadronic energy,
which experimentally is always the case. Furthermore, hadronization corrections
certainly produce some smearing on the partonic final states because
of parton recombination.
In other words, non-perturbative mechanisms wash out this infrared divergence, which
therefore does not present any problem of principle.
As we are going to show, however, perturbation theory ``saves itself'' and
no mechanism outside perturbation theory is needed to have a consistent prediction:
resummation of the infrared logarithms in (\ref{logsintro})
to all orders completely eliminates the
singularity, as in the cases of the usual infrared divergencies \cite{sudshould}.
Since large logarithms occur for $E_X \sim m_b/2$, we have that the hard scale
is given for this spectrum by the beauty mass,
\begin{equation}
~~~~~~~~~~Q \, = \, m_b~~~~~~~~~~~~~{\rm (hadron~energy~spectrum)},
\end{equation}
just like in radiative decays. This equality is noticeable,
as it comes from completely independent kinematics with respect to the one
in (\ref{bsgamma}).
There is therefore a pure short-distance relation between the hadron mass
distribution in (\ref{bsgamma}) and the hadron energy distribution in
(\ref{semilep}). This property remains true when we consider non-perturbative
Fermi motion effects, which are factorized by the well-known structure function
of the heavy flavors, also called the shape function.

This paper is organized as follows:

In sec.~(\ref{triplodiff})
we presents the results for the resummed triple-differential
distribution, which is the most general distribution and the starting point
of our analysis;

In sec.~(\ref{secthresum}) we review the theory of threshold resummation
in heavy flavor decays, giving explicit formulas for the QCD form factor
in next-to-next-to-leading order (NNLO).
The transformation to Mellin space in order to solve the kinematical constraints
for multiple soft emission is discussed, together with the inverse
transform to the original momentum space;

In sec.~(\ref{hadrvars}) we derive the double distribution in the hadronic
energy and in the ratio (hadronic mass)/(hadronic energy), which are
the most convenient variables for threshold resummation
(these are the variables $w$ and $u$ defined there).
The distribution in any hadronic variable can be obtained from
this distribution by integration;

In sec.~(\ref{hadspec}) we present the results for the resummed
hadron energy spectrum in next-to-leading order, whose main
physical properties have already been anticipated here.
We also compute the average hadronic energy to first order and
compare with the radiative decay.
The hadron energy spectrum with an upper cutoff on the hadron mass,
which is the easiest thing to measure in experiments, is derived
in leading order;

In sec.~(\ref{sec2energies}) we derive the double distribution in the hadron and
electron energies, i.e. in the two independent energies.
A peculiarity of this spectrum is that it is characterized by the presence
of two different series of large logarithms, which are factorized
by two different QCD form factors.
Another peculiarity is that this double differential distribution
contains partially-integrated QCD form factors instead of
differential ones. That implies that the infrared singularities
occurring in this distribution are integrable, as in the case of
the Sudakov shoulder which we have discussed before;

Finally, in sec.~(\ref{concl}) we present our conclusions together
with a discussion about natural developments.

\section{Triple differential distribution}
\label{triplodiff}

The triple differential distribution in the decay (\ref{semilep})
is the starting point of our analysis.
It has a resummed expression of the form \cite{me}:\footnote{
We have normalized the distribution to the radiatively-corrected
total semileptonic width
$\Gamma=\Gamma_0\left[1 + \alpha C_F/\pi\left( 25/8 - \pi^2/2 \right) +O(\alpha^2) \right]$
and not to the Born width $\Gamma^{(0)}$, as originally done in \cite{me}.
We consider it to be a better choice because $\Gamma$, unlike $\Gamma^{(0)}$,
is a physical quantity, directly measurable in the experiments and
we are not interested in the prediction of total rates, but only
in how a given rate distributes among different hadronic channels.}
\begin{equation}
\label{tripla}
\frac{1}{\Gamma}\frac{d^3\Gamma}{dx du dw} \,=\, C\left[x,w;\alpha(w\,m_b)\right]\,
\sigma\left[u;\alpha(w\,m_b)\right] \, + \, d\left[x,u,w;\alpha(w\,m_b)\right],
\end{equation}
where we have defined the following kinematical variables:
\begin{equation}
w \,=\, \frac{2 E_X}{m_b}~~~~~~~~(0\le w \le 2),~~~~~~~~~~~~~~~~~~~~~
x \,=\, \frac{2 E_l}{m_b}~~~~~~~(0\le x \le 1)
\end{equation}
and
\footnote{Note that a similar variable simplifies two-loop computations with
heavy quarks \cite{uanalog}.}
\begin{equation}
u \, = \, \frac{E_X - \sqrt{E_X^2 - m_X^2} }{E_X + \sqrt{E_X^2 - m_X^2} }
~~~~~~~~~~(0\le u \le 1).
\end{equation}
It is convenient to express $u$ as:
\begin{equation}
\label{defu}
u \, = \, \frac{1\, - \, \sqrt{1-4y}}{1\, + \, \sqrt{1-4y}},
\end{equation}
with
\begin{equation}
\label{defy}
y \, = \, \frac{m_X^2}{4 E_X^2}~~~~~~~~~~~~~~~(0\le y\le 1/4).
\end{equation}
The inverse formula of (\ref{defu}) reads:
\begin{equation}
y\,=\,\frac{u}{(1+u)^2}.
\end{equation}
The functions entering the r.h.s. of eq.~(\ref{tripla}) are:
\begin{itemize}
\item
$C\left[x,w;\alpha(w\,m_b)\right]$, a short-distance, process-dependent
coefficient function, whose explicit expression will be given later.
It depends on two independent energies $x$ and $w$ and on the QCD coupling $\alpha$;
\item
$\sigma\left[u;\alpha(w\,m_b)\right]$, a
process-independent, long-distance dominated, QCD form factor. It
factorizes the threshold logarithms appearing in the perturbative expansion.
At order $\alpha$:
\begin{equation}
\label{sigma1}
\sigma(u;\alpha) \, = \, \delta(u) \, - \, \frac{C_F \,\alpha}{\pi}
\left(\frac{\log u}{u}\right)_+
                \, - \, \frac{ 7 \, C_F\, \alpha}{4\,\pi} \left(\frac{1}{u}\right)_+
                \, + \, O(\alpha^2),
\end{equation}
where $C_F$ is the Casimir of the fundamental representation of
$SU(3)_c$, $C_F=(N_c^2-1)/(2N_c)$ with $N_c=3$ (the number of
colors) and the plus distributions are defined as usual as:
\begin{equation}
P(u)_+ \,=\, P(u) - \delta(u) \int_0^1 d u' P(u').
\end{equation}
The action on a test function $f(u)$ is therefore:
\begin{equation}
\int_0^1 du \, P(u)_+ \, f(u) \,=\, \int_0^1 du \, P(u)\,
[f(u) \, - \, f(0)].
\end{equation}
The plus-distributions are sometimes called star-distributions and can
also be defined as limits of ordinary functions as:
\begin{eqnarray}
P(u)_+ &=& \lim_{\epsilon\rightarrow 0^+}
\Big[
\theta(u-\epsilon) \, P(u)
\, - \, \delta(u)\, \int_{\epsilon}^{1}du' \, P(u')
\Big]
\nonumber\\
&=&  \lim_{\epsilon\rightarrow 0^+}
\Big[
\theta(u-\epsilon) \, P(u)
     \, - \, \delta(u-\epsilon)\, \int_{\epsilon}^{1}du' \, P(u')
\Big]
\nonumber\\
&=& \lim_{\epsilon\rightarrow 0^+} - \, \frac{d}{du}
\Big[
\theta(u-\epsilon)\,\int_u^1 du' \, P(u')
\Big].
\end{eqnarray}
An important property of the plus-distributions is that their integral on the
unit interval vanishes:
\begin{equation}
\int_0^1 \, P(u)_+ \, d u \, = \, 0.
\end{equation}
We have assumed a minimal factorization scheme in eq.~(\ref{sigma1}),
in which only terms containing plus-distributions are included in the form
factor.
The resummation of the logarithmically enhanced terms in $\sigma$ to
all orders in perturbation theory will be discussed in the next section;
\item
$d\left(x,u,w;\,\alpha\right)$ is a short-distance, process-dependent, remainder
function, not containing large logarithms. Formally, it can have at most
an integrable singularity for $u\rightarrow + \,0$, i.e. we require that:
\begin{equation}
\lim_{u\rightarrow + 0}
\int_0^u d u' d(x,w,u';\alpha) \, = \, 0.
\end{equation}
This term is added to $C\cdot\sigma$ in order to correctly describe the
region $u \sim O(1)$ and to reproduce the total rate.
It depends on all the kinematical variables $x,\,w$ and $u$ and the explicit expression
will be given later.
\end{itemize}
Eq.~(\ref{tripla}) is a generalization of the threshold resummation formula for the
radiative decay in (\ref{bsgamma}) \cite{me,ucg}:
\begin{equation}
\label{radsum}
\frac{1}{\Gamma_R}\frac{d\Gamma_R}{d t_s} \,=\, C_R\left[\alpha(w\,m_b)\right]\,
\sigma\left[t_s;\,\alpha(w\,m_b)\right] \, + \, d_R\left[t_s;\,\alpha(w\,m_b)\right],
\end{equation}
where\footnote{
The relation with the photon energy $x_{\gamma}=2E_{\gamma}/m_b$ is $t_s=1-x_{\gamma}$.}
\begin{equation}
t_s \, = \, \frac{m_{X_s}^2}{m_b^2}.
\end{equation}
In this simpler case, the coefficient function $C_R(\alpha)$
does not depend on any kinematical variable but only on the QCD coupling
$\alpha$ and has an expansion of the form\footnote{
We perform expansions in powers of $\alpha$, while the traditional expansion
is in powers of $\alpha/(2\pi)$.}:
\begin{equation}
C_R(\alpha) \, = \, 1 \, + \, \alpha \, C_R^{(1)} \, + \, \alpha^2 \, C_R^{(2)} \,
+ \, O(\alpha^3),
\end{equation}
where $C_R^{(i)}$ are numerical coefficients.
Basically, going from the 2-body decay (\ref{bsgamma}) to the 3-body decay
(\ref{semilep}), the coefficient function acquires a dependence on the additional
kinematical variables, namely two energies.
The remainder function in eq.~(\ref{radsum}) depends on the (unique) variable $t_s$
and has an expansion of the form:
\begin{equation}
d_R(t_s;\,\alpha) \, = \, \alpha \, d_R^{(1)}(t_s) \, + \,
 \alpha^2 \, d_R^{(2)}(t_s) \, + \, O(\alpha^3).
\end{equation}
The main point is that the QCD form factor $\sigma$ in the same in both
distributions (\ref{tripla}) and (\ref{radsum}), explicitly showing universality
of long-distance effects
in the two different decays. By universality we mean that we have the same function,
evaluated at the argument $u$ in the semileptonic case and at
$t_s$ in the radiative decay.
This property is not explicit in the original formulation
\cite{akhouri}, in which the form factors differ in subleading order
(see next section).

Since, as shown in the introduction, $w\sim 1$ in the radiative decay, we can
make everywhere in eq.~(\ref{radsum}) the replacement
\begin{equation}
\label{setwone}
~~~~~~~~~\alpha(w\,m_b) \, \rightarrow \, \alpha(m_b)
~~~~~~~~~~~~~~{\rm(radiative~case~only)},
\end{equation}
to obtain:
\begin{equation}
\label{radsum2}
\frac{1}{\Gamma_R}\frac{d\Gamma_R}{d t_s} \,=\, C_R\left[\alpha(m_b)\right]\,
\sigma\left[t_s;\,\alpha(m_b)\right] \, + \, d_R\left[t_s;\,\alpha(m_b)\right].
\end{equation}
The distribution contains now a constant coupling, independent on the kinematics
$\alpha(m_b)\simeq 0.22$.
The replacement (\ref{setwone}) cannot be done in the semileptonic case.

\noindent
In \cite{me} the triple differential distribution was originally given
in terms of the variable $y$ instead of $u$, with the latter $u=1-\xi$
being introduced in \cite{ucg}.
The variables $u$ and $y$ coincide in the threshold region in leading twist,
i.e. at leading order in $u$ in the expansion for $u\rightarrow 0$,
as $y\,= \, u \, + \, O(u^2)$.
Going from the variable $y$ to the variable $u$
only modifies the remainder function.
The advantages of $u$ over $y$ are both technical and physical:
\begin{itemize}
\item
$u$ has, unlike $y$, unitary range;
\item
when we impose the kinematical relation between hadronic energy $E_{X_s}$ and
hadronic mass $m_{X_s}$ of the radiative decay (\ref{bsgamma}),
$u$ exactly equals $t_s$:
\begin{equation}
u|_{E_{X_s} = m_b/2(1+m_{X_s}^2/m_b^2)} \, = \, t_s.
\end{equation}
This property suggests that some higher-twist effects may cancel in
taking proper ratios of radiative and semileptonic spectra.
\end{itemize}
Let us now give the explicit expression of the coefficient function
in the semileptonic case:
\begin{equation}
\label{Cexpansion}
C(\overline{x},w;\,\alpha) \, = \,
C^{(0)}(\overline{x},w) \, + \, \alpha \, C^{(1)}(\overline{x},w)
\, + \, \alpha^2 \,C^{(2)}(\overline{x},w) \, + \, O(\alpha^3),
\end{equation}
where
\begin{eqnarray}
C^{(0)}(\overline{x},w) &=& 12 (w-\overline{x})(1+\overline{x}-w);
\\
C^{(1)}(\overline{x},w) &=& 12 \frac{C_F}{\pi} (w-\overline{x}) \Bigg\{
(1+\overline{x}-w)
\left[
{\rm Li}_2(w) + \log w \log(1-w)
-\frac{3}{2}\log w - \frac{w \log w}{2(1-w)} - \frac{35}{8}
\right] +
\nonumber\\
&&~~~~~~~~~~~~~~~~~~ + \, \frac{\overline{x}\,\log w}{2(1-w)}
\Bigg\}
\end{eqnarray}
with $\overline{x}=1-x$ \footnote{ To avoid spurious imaginary
parts for $w>1$ one can use the relation $ {\rm Li}_2(w) = - {\rm
Li}_2(1-w) - \log w\log(1-w) +\pi^2/6$.}.
Note that the coefficient function contains the overall factor
$w-\overline{x}=\overline{x}_{\nu}$, which vanishes linearly at
the endpoint of the neutrino spectrum. We have defined
$\overline{x}_{\nu} = 1-x_{\nu}$ and $x_{\nu} = 2E_{\nu}/m_b$.

\noindent
Unlike the coefficient function, the remainder function
$d\left(x,u,w;\alpha\right)$ has an expansion starting at $O(\alpha)$:
\begin{equation}
\label{dexpansion}
d\left(x,w,u;\,\alpha\right) \, = \, \alpha \,  d^{(1)}\left(x,w,u\right)
\,+\, \alpha^2 \, d^{(2)}\left(x,w,u\right) \, + \, O(\alpha^3).
\end{equation}
Omitting the overall factor $C_F/\pi$, we obtain:
\footnote{
The $O(\alpha)$ function is obtained from that one given in \cite{me}
$d^{(1)}_{old}\left(x,w,y\right)$ in terms of the variables
$z = 1 - y$ and $\zeta = 1 - 4 y$,
by using a relation extending eq.~(23) of \cite{ucg}:
\begin{equation}
d^{(1)}\left(x,w,u\right) \, = \, d^{(1)}_{old}\left(x,w,y(u)\right) \, \frac{dy}{du}(u)
\,+\, C^{(0)}(x,w) \frac{C_F}{\pi}
\left[
\frac{\log u + 7/4}{u} \, - \, \frac{\log y(u) + 7/4}{y(u)} \, \frac{dy}{du}(u)
\right].
\end{equation}
This function can also be obtained with a direct matching
with the $O(\alpha)$ triple differential distribution computed in \cite{ndf}
after a change of variable (see the end of this section for a discussion
about matching).}
\begin{eqnarray}
d^{(1)}(\overline{x},w,u) &=&
\frac{-3\,w^4\,\left( 24 + 3\,w -
       8\,\overline{x} \right) }{4\,
     {\left( 1 + u \right) }^5} +
  \frac{9\,w^4\,\left( 24 + 3\,w -
       8\,\overline{x} \right) }{8\,
     {\left( 1 + u \right) }^4} +
\nonumber\\
&-& \frac{9\,\left( -12 + w \right) \,
     {\left( -2 + w \right) }^2\,
     {\left( w - 2\,\overline{x} \right) }^2}
     {16\,{\left( 1 - u \right) }^3} +
  \frac{9\,\left( -12 + w \right) \,
     {\left( -2 + w \right) }^2\,
     {\left( w - 2\,\overline{x} \right) }^2}
     {32\,{\left( 1 - u \right) }^2} +
\nonumber\\
&-& \frac{3\,w^2\,\left( 32 - 47\,w - 8\,w^2 +
       16\,\overline{x} +
       20\,w\,\overline{x} +
       w^2\,\overline{x} +
       8\,{\overline{x}}^2 -
       3\,w\,{\overline{x}}^2 \right) }{8\,
     {\left( 1 + u \right) }^2} +
\nonumber\\
&-& \frac{3\,w^2\,\left( -64 + 94\,w + 40\,w^2 +
       3\,w^3 - 32\,\overline{x} -
       40\,w\,\overline{x} -
       10\,w^2\,\overline{x} -
       16\,{\overline{x}}^2 +
       6\,w\,{\overline{x}}^2 \right) }{8\,
     {\left( 1 + u \right) }^3} +
\nonumber\\
&+& \frac{3 }{64 \, \left( 1 + u \right) }
\,\Big( 640\,w - 368\,w^2 - 200\,w^3 -
       16\,w^4 + 3\,w^5 - 384\,\overline{x} +
       320\,w\,\overline{x} +
       528\,w^2\,\overline{x} +
\nonumber\\
&+&    112\,w^3\,\overline{x} -
       16\,w^4\,\overline{x} -
       256\,{\overline{x}}^2 -
       48\,w\,{\overline{x}}^2 -
       224\,w^2\,{\overline{x}}^2 +
       24\,w^3\,{\overline{x}}^2 \Big) +
\nonumber\\
&+&  \frac{3}{64\,\left( 1 - u \right) }
\,\Big( -256\,w + 528\,w^2 - 200\,w^3 -
       16\,w^4 + 3\,w^5 + 512\,\overline{x} -
       1472\,w\,\overline{x} +
\nonumber\\
&+&    528\,w^2\,\overline{x} +
       112\,w^3\,\overline{x} -
       16\,w^4\,\overline{x} +
       640\,{\overline{x}}^2 -
       48\,w\,{\overline{x}}^2 -
       224\,w^2\,{\overline{x}}^2 +
       24\,w^3\,{\overline{x}}^2 \Big) +
\nonumber\\
&-& \frac{9\,w^5\,\log \, u}
   {4\,{\left( 1 + u \right) }^6} +
  \frac{9\,w^5\,\log \, u}
   {2\,{\left( 1 + u \right) }^5} -
  \frac{9\,\left( -12 + w \right) \,
     {\left( -2 + w \right) }^2\,
     {\left( w - 2\,\overline{x} \right) }^2\,
     \log \, u}{16\,{\left( 1 - u \right) }^4} +
\nonumber\\
&+& \frac{9\,\left( -12 + w \right) \,
     {\left( -2 + w \right) }^2\,
     {\left( w - 2\,\overline{x} \right) }^2\,
     \log \, u}{16\,{\left( 1 - u \right) }^3} +
\nonumber\\
&+& \frac{3\,w^3\,\left( -10 + 16\,w + w^2 +
       8\,\overline{x} - 2\,w\,\overline{x} -
       2\,{\overline{x}}^2 \right) \,\log \, u}
     {8\,{\left( 1 + u \right) }^3} +
\nonumber\\
&-& \frac{3\,w^3\,\left( -10 + 16\,w + 7\,w^2 +
       8\,\overline{x} - 2\,w\,\overline{x} -
       2\,{\overline{x}}^2 \right) \,\log \, u}
     {8\,{\left( 1 + u \right) }^4} +
\nonumber\\
&-& \frac{3\log \, u}{64\,{\left( 1 + u \right) }^2}
\,w\,\Big( -144\,w + 208\,w^2 +
       16\,w^3 + w^4 - 64\,\overline{x} -
       80\,w\,\overline{x} -
       16\,w^2\,\overline{x} +
\nonumber\\
&-&    8\,w^3\,\overline{x} +
       48\,{\overline{x}}^2 -
       96\,w\,{\overline{x}}^2 +
       16\,w^2\,{\overline{x}}^2 \Big) \, +
\nonumber\\
&+& \frac{3 \log \, u}{64\,{\left( 1 - u \right) }^2}
\,\Big( -256\,w + 624\,w^2 - 304\,w^3 +
       16\,w^4 + w^5 + 512\,\overline{x} -
       1856\,w\,\overline{x} +
\nonumber\\
&+&    944\,w^2\,\overline{x} -
       16\,w^3\,\overline{x} -
       8\,w^4\,\overline{x} +
       1024\,{\overline{x}}^2 -
       464\,w\,{\overline{x}}^2 -
       96\,w^2\,{\overline{x}}^2 +
       16\,w^3\,{\overline{x}}^2 \Big).
\end{eqnarray}
The remainder function is a combination of rational functions of $u$
multiplied in some cases by $\log u$, with coefficients given by polynomials
in $w$ and $\overline{x}$.

The main point about the semileptonic decay (\ref{semilep}) is that it has --- unlike the
radiative decay (\ref{bsgamma}) --- $q^2\ne 0$ and consequently the form factor
depends not only on $u$ but also on the hadronic energy $w$ through the
running coupling:
\begin{equation}
\sigma\,=\,\sigma[u;\,\alpha(w\,m_b)].
\end{equation}
The form factor is therefore a function of two variables.

We work in next-to-leading order (NLO),
in which only the $O(\alpha)$ corrections
to the coefficient function and remainder function are retained
(see next section).
Since the difference between $\alpha(w\,m_b)$ and $\alpha(m_b)$ is
$O(\alpha^2)$, we can set $w=1$ in the argument of the coupling entering
the coefficient function and the remainder function.
We then obtain the simpler expression:
\begin{equation}
\label{triplefin}
\frac{1}{\Gamma}\frac{d^3\Gamma}{dx du dw} \, = \,
C\left[x,w;\alpha(m_b)\right]\,
\sigma\left[u;\alpha(w\,m_b)\right] \, + \,
d\left[x,u,w;\alpha(m_b)\right]~~~~~~~~~~~~~~~{\rm (NLO)}.
\end{equation}
Note that we cannot set $w=1$ in the coupling entering the form factor,
because in the latter case $\alpha$ is multiplied by large logarithms,
which ``amplify'' $O(\alpha^2)$ differences in the couplings
(see next section).

\noindent
Let us make a few remarks about the final result of this section,
eq.~(\ref{triplefin}):
\begin{itemize}
\item
it describes semi-inclusive decays, in which the internal structure of the
hadronic final states is not observed, but only the total mass and energy
are measured.
Less inclusive quantities, such as for instance the energy distribution of the final
$up$ quark (i.e. the fragmentation function of the $up$ quark),
cannot be computed in this framework;
\item
it constitutes an improvement of the fixed-order $O(\alpha)$
result in all the cases in which there are large threshold logarithms.
In all the other cases, where there are no threshold logarithms,
such as for example the dilepton mass distribution \cite{dileptonspec}, there
is not any advantage of the resummed formula over the fixed-order one.
\end{itemize}
In the next sections we integrate the resummed triple-differential distribution
to obtain double and single (resummed) spectra.
There are two methods to accomplish this task which are completely
equivalent:
\begin{enumerate}
\item
The first method involves the direct integration of the complete triple-differential
distribution. Schematically:
\begin{equation}
\label{schematic}
{\rm(spectrum)} \, = \, \int C \cdot \sigma \, + \, \int d.
\end{equation}
Large logarithms come only from the first term on the r.h.s. of (\ref{schematic}),
while non-logarithmic, ``small'' terms come both from the first and the second term.
To obtain a factorized form for the spectrum analogous to the one for the triple-distribution,
in which the remainder function collects {\it all} the small terms, one rearranges
the r.h.s. of (\ref{schematic}): the small terms coming from the integration of $C\cdot\sigma$
are put in the remainder function;
\item
In the second method, one integrates the block $C\cdot\sigma$ only and drops the small terms
coming from the integration.
The remainder function is obtained by expanding the resummed expression in powers
of $\alpha$ and comparing with the fixed-order spectrum.
\end{enumerate}

\section{Threshold Resummation}
\label{secthresum}

It is convenient to define  the partially integrated or cumulative form factor
$\Sigma(u,\alpha)$:
\begin{equation}
\Sigma(u;\,\alpha) \,=\,\int_0^u d u^\prime \, \sigma(u^\prime;\,\alpha).
\end{equation}
Performing the integrations, one obtains for the $O(\alpha)$ form
factor:
\begin{equation}
\label{Sigmaalpha}
\Sigma\left(u;\,\alpha\right) \,=\, 1 \, - \, \frac{C_F \,\alpha}{2\,\pi} \, L^2
\, + \, \frac{ 7\,C_F\, \alpha}{4\,\pi}\, L \, + \, O(\alpha^2),
\end{equation}
where
\begin{equation}
L \, = \, \log\frac{1}{u}.
\end{equation}
$\Sigma$ contains a double logarithm coming from the overlap
of the soft and the collinear region and a single logarithm
of soft or collinear origin.
The normalization condition reads:
\begin{equation}
\Sigma(1;\,\alpha) \,=\, \int_0^1 d u \, \sigma(u;\alpha) \,=\, 1.
\end{equation}
As already noted, we have assumed a minimal factorization scheme, in which
only logarithms and not constants or other functions are contained
in the form factor.
The expression of the partially integrated form
factor $\Sigma$ is technically simpler than the one for the differential
form factor $\sigma$, as it involves ordinary functions instead of
generalized ones. Furthermore, in experiments one always measures some
integral of $\sigma$ around a central $u$ value because of the binning.

In the limit $u\rightarrow 0^+$, no final states are included in the distribution
and therefore one expects, on physical grounds, that
\begin{equation}
\label{shouldbeso}
\lim_{u\rightarrow 0^+} \, \Sigma(u;\,\alpha) \, = \, 0.
\end{equation}
The $O(\alpha)$ expression (\ref{Sigmaalpha}) does not have this
limit and it is actually divergent to $-\infty$ --- a completely
un-physical result. In general, a truncated expansion in powers of
$\alpha$ is divergent for $u\rightarrow 0^+$, because the
coefficients diverge in this limit. Therefore, one  has to resum
the infrared logarithms, i.e. the terms of the form $\alpha^n \,
L^k$, to all orders in perturbation theory. In higher orders,
$\Sigma$ contains at most two logarithms for each power of
$\alpha$, one of soft origin and another one of collinear origin.
Its general expression is then:
\begin{equation}
\label{Sigmaexp}
\Sigma\left(L,\alpha\right) \, = \, 1 \, +\, \sum_{n=1}^{\infty}
\sum_{k=1}^{2n} \Sigma_{n k} \, \alpha^n \, L^k,
\end{equation}
where $\Sigma_{n k}$ are numerical coefficients.
At present, a complete resummation of all the logarithmically-enhanced
terms on the r.h.s. of eq.~(\ref{Sigmaexp}) is not feasible in QCD: one has to resort to
approximate schemes. The most crude approximation consists of picking
up the most singular term for $u\rightarrow 0^+$ for each power of
$\alpha$, i.e. all the terms of the form:
\begin{equation}
~~~~~~~~~~~~~~~\alpha^n\,L^{2n}~~~~~~~~~~~~~~~~~~~~~~~~~~
{\rm (double~logarithmic~approximation).}
\end{equation}
In this approximation, we can neglect running coupling effects and effects related
to the kinematical constraints:
higher orders simply exponentiate the $O(\alpha)$ double logarithm and one
obtains
\begin{equation}
\label{resumDLA}
~~~~~~~~~~~~~~~~
\Sigma\left(u;\,\alpha\right) \,=\, e^{ - \, C_F\,\alpha/(2\pi) \, L^2 }
~~~~~~~~~~~~~~~~~~~~~~~~~~~~~
{\rm (double~logarithmic~approximation).}
\end{equation}
Let us note that the resummed expression (\ref{resumDLA}),
unlike the fixed-order one (\ref{Sigmaalpha}), does satisfy the
condition (\ref{shouldbeso}). The exponent in the resummed form factor involves a single term,
$- \, C_F\,\alpha/(2\pi) \, L^2$,
and has therefore a simpler form than the form factor itself. This remains true when
more accurate resummation schemes are constructed, so it is convenient to define $G$
as:
\begin{equation}
\Sigma \, = \, e^G.
\end{equation}
It can be shown that the expansion for the function $G$ is of the form \cite{backtox}:
\begin{equation}
\label{expG}
G(L;\alpha) \,=\, \sum_{n=1}^{\infty} \sum_{k=1}^{n+1} G_{n k}\, \alpha^n \, L^k,
\end{equation}
where $G_{n k}$ are numerical coefficients.
Let us note that the sum over $k$ extends up to $n+1$ in (\ref{expG}), while it
extends up to $2n$ in the form factor in eq.~(\ref{Sigmaexp}).
This property is a generalization of the simple exponentiation of the $O(\alpha)$
logarithms which holds in QED and is called generalized exponentiation.
In general, this property holds for quantities analogous to the semi-inclusive
form factors, in which the gluon radiation is not directly observed.
One sums therefore over all possible final states coming from the evolution of the
emitted gluons (inclusive gluon decay quantities).
The property expressed by eq.~(\ref{expG}) does not hold for quantities in which
gluon radiation is observed directly, as for example in parton
multiplicities, where different evolutions of gluon jets give rise to
different multiplicities.

\subsection{$N$-space}

A systematic resummation is consistently done in $N$-moment space or Mellin space,
in which kinematical constraints are factorized in the soft limit and are easily
integrated over \cite{catani}.
One considers the Mellin transform of the form factor $\sigma(u;\,\alpha)$:
\begin{equation}
\sigma_N(\alpha) \, \equiv \, \int_0^1 du \, (1-u)^{N-1} \, \sigma(u;\,\alpha).
\end{equation}
The threshold region is studied in moment space by taking the limit
$N\rightarrow\infty$, because for large $N$ the integral above takes
contributions mainly from the region $u\ll 1$.
For example, the Mellin transform  of the spectrum in eq.~(\ref{radsum2})
is of the form
\begin{equation}
\label{mellindist}
\int_0^1 (1-t_s)^{N-1}
\frac{1}{\Gamma_R}\frac{d\Gamma_R}{d t_s} \, dt_s\,=\, C_R(\alpha)\,
\sigma_N(\alpha) \, + \, d_{R,N}(\alpha),
\end{equation}
where
\begin{equation}
d_{R,N}(\alpha) \, \rightarrow \, 0~~~~~~~~~~~~~~{\rm for}~N\,\rightarrow\,\infty.
\end{equation}
The total rate in Mellin space is obtained by taking $N=1$.

It can be shown \cite{sterman,cattren1,cattren2}
that the form factor in $N$-space has the following exponential structure:
\begin{equation}
\label{thresum}
\sigma_N(\alpha) \, = \, e^{ G_N(\alpha) },
\end{equation}
where
\begin{equation}
\label{expresum}
G_N(\alpha) \, = \,
\int_0^1 dz \frac{ z^{N-1} - 1 }{1-z}
\left\{
\int_{Q^2(1-z)^2}^{Q^2(1-z)} \frac{dk_t^2}{k_t^2}
A\left[\alpha(k_t^2)\right]
\, + \, B\left[\alpha(Q^2(1-z))\right]
\, + \, D\left[\alpha(Q^2(1-z)^2)\right]
\right\}.
\end{equation}
Let us note that a prescription has to be assigned to this formula
since it involves integrations over the Landau pole \cite{model}.
The functions entering the resummation formula have a standard
fixed-order expansion, with numerical coefficients:
\begin{eqnarray}
A(\alpha) &=& \sum_{n=1}^{\infty} A_n\,\alpha^n \, = \, A_1 \alpha +  A_2 \alpha^2  + A_3 \alpha^3
+ A_4 \alpha^4 + \cdots
\\
B(\alpha) &=& \sum_{n=1}^{\infty} B_n\,\alpha^n \, = \, B_1 \alpha +  B_2 \alpha^2 + B_3 \alpha^3  + \cdots
\\
D(\alpha) &=& \sum_{n=1}^{\infty} D_n\,\alpha^n \, = \, D_1 \alpha +  D_2 \alpha^2 + D_3 \alpha^3  + \cdots
\end{eqnarray}
The known values for the resummation constants read:
\begin{eqnarray}
A_1 &=& \frac{C_F}{\pi};
\\
A_2 &=&  \frac{C_F}{\pi^2} \left[ C_A\left( \frac{67}{36} - \frac{z(2)}{2} \right)
- \frac{5}{18} n_f \right];
\\
A_3 &=& \frac{C_F}{\pi^3}
\Bigg[
C_A^2\Big(
\,\frac{245}{96} + \frac{11}{24} z(3) - \frac{67}{36} \, z(2) + \frac{11}{8} \, z(4)
\Big)
\, - \, C_A \, n_f \Big( \frac{209}{432} + \frac{7}{12} z(3) - \frac{5}{18}\,z(2) \Big) +
\nonumber\\
&&~~~~~~ - \, C_F \, n_f \Big( \frac{55}{96} - \frac{z(3)}{2}  \Big)
- \frac{n_f^2}{108}
\Bigg];
\\
B_1 &=&  - \frac{3}{4} \frac{C_F}{\pi};
\\
B_2 & = & \frac{C_F}{\pi^2}
\left[
C_A \left( - \frac{3155}{864} + \frac{11}{12}\,z(2) + \frac{5}{2} \, z(3) \right)
- C_F \left( \frac{3}{32} + \frac{3}{2} z(3) - \frac{3}{4} \, z(2) \right)
+ n_f \left( \frac{247}{432} - \frac{z(2)}{6} \right)
\right];
\\
D_1 &=&  - \frac{C_F}{\pi};
\\
D_2 &=& \frac{C_F}{\pi^2} \left[
C_A \left( \frac{55}{108} - \frac{9}{4} z(3) + \frac{z(2)}{2} \right)
+ \frac{n_f}{54}\right],
\end{eqnarray}
where $C_A=N_c=3$ is the Casimir of the adjoint representation.
The coefficients $A_1$, $B_1$ and $D_1$ are renormaliza\-tion-scheme independent, as they can be
obtained from tree-level amplitudes with one-gluon emission (see later).
The higher-order coefficients are instead renormalization-scheme dependent
and are given in the $\overline{MS}$ scheme for the coupling constant
\footnote{A discussion about the scheme dependence of the higher
order coefficients $A_2,\,B_2,$ etc. on the coupling constant
can be found in \cite{catweb}.}.

To this approximation, the first three orders of the $\beta$-function are
also needed
\cite{beta2,beta3}:
\begin{eqnarray}
\beta_0&=& \frac{1}{4\pi} \Big[ \frac{11}{3} C_A - \frac{2}{3} n_f \Big];
\\
\beta_1&=& \frac{1}{24\pi^2} \Big[ 17 \, C_A^{\,2} - \Big(5 \, C_A + 3 \, C_F\Big) n_f \Big];
\\
\beta_2 &=& \frac{1}{64 \pi^3}
\Big[
\frac{2857}{54} C_A^{\,3}
- \left(
\frac{1415}{54}C_A^{\,2} + \frac{205}{18} C_A C_F - C_F^{\,2}
\right) n_f
+\left(
\frac{79}{54} C_A + \frac{11}{9} C_F
\right) n_f^2
\Big].
\end{eqnarray}
As is well known, $\beta_0$ and $\beta_1$ are renormalization-scheme independent,
while $\beta_2$ is not and has been given in the $\overline{MS}$ scheme.
We define the $\beta$-function with an overall minus sign:
\begin{equation}
\frac{d\alpha}{d\log\mu^2}\,=\,-\,\beta(\alpha)\,
= \,-\,\beta_0\,\alpha^2 \,-\,\beta_1\,\alpha^3\,-\,\beta_2\,\alpha^4 \, - \, \cdots.
\end{equation}
The running coupling reads:
\begin{equation}
\alpha(\mu) \, = \, \frac{1}{\beta_0\, \log \mu^2/\Lambda^2}
\, - \, \frac{\beta_1}{\beta_0^3} \frac{\log\left(\log \mu^2/\Lambda^2\right)}{\log^2 \mu^2/\Lambda^2}
\, + \, \frac{\beta_1^2}{\beta_0^5}
\frac{\log^2\left(\log \mu^2/\Lambda^2\right) \, - \, \log\left(\log \mu^2/\Lambda^2\right) \, - \, 1}
{\log^3 \mu^2/\Lambda^2}
\, + \, \frac{\beta_2}{\beta_0^4} \frac{ 1 }{\log^3 \mu^2/\Lambda^2}.
\end{equation}
The functions $A(\alpha)$, $B(\alpha)$ and $D(\alpha)$ have the following
physical interpretation (see for example \cite{proclecce,conlucastud}):
\begin{itemize}
\item
The function $A(\alpha)$ involves a double integration over the transverse momentum $k_t$ and
the energy $\omega$ of the emitted gluon and represents emissions
at small angle and at small energy from the light quark.
The leading term $A_1$ is the coefficient of that piece of the matrix element squared
for one real gluon emission, which is singular in the small angle and small energy
limit:
\begin{equation}
\label{doublerepr}
A_1 \, \alpha \, \frac{d\omega}{\omega} \, \frac{d\theta^2}{\theta^2}
\, \cong \, A_1 \, \alpha \, \frac{d\omega}{\omega} \, \frac{d k_t^2}{k_t^2},
\end{equation}
where $k_t\, \simeq \, \omega\,\theta$ is the transverse momentum of the gluon.
In (\ref{doublerepr}) we have given the representation of the integral both in the angle $\theta$
and in the transverse momentum $k_t$.
The subleading coefficients $A_2$, $A_3$, etc. represent corrections to the basic
double-logarithmic emission.
The function $A(\alpha)$ ``counts'' the number of light quark jets in different
processes, i.e.
we can write
\begin{equation}
A^{(P)}(\alpha) \, = \, n_q \, A(\alpha),
\end{equation}
where $n_q$ is the number of primary light quarks in the process $P$.
For example, in $e^+e^-$ annihilation into hadrons $n_q=2$, while in
the heavy flavor decays (\ref{seminclusive}) $n_q=1$.
Since soft gluons only couple to the four-momentum of their emitters
and not to their spin, the function $A_g(\alpha)$ for gluon jets is obtained
from the quark one $A(\alpha)$ simply taking into account the change in the color
charge, i.e. multiplying by $C_A/C_F$ \cite{demilio};
\item
the function $B(\alpha)$ represents emissions at small angle with a large
energy from the light quark.
$B_1$ is the coefficient of that piece of the matrix element squared
which is singular in the small angle limit:
\begin{equation}
\label{brepr}
B_1 \, \alpha \, d\omega  \, \frac{d\theta^2}{\theta^2} \, \cong \,
B_1 \, \alpha \, d\omega  \, \frac{d k_t^2}{k_t^2}.
\end{equation}
The non logarithmic integration over the gluon energy $\omega$ has been done and does not
appear explicitly in eq.~(\ref{expresum}); the integration over the angle $\theta$ or
the transverse momentum $k_t$ is rewritten as an integral over $z$.
The function $B(\alpha)$ counts the number of final-quark jets, i.e.
\begin{equation}
B^{(P)}(\alpha) \, = \, n_l \, B(\alpha),
\end{equation}
where $n_l$ is the number of primary final quarks in the process $P$.
For example in $e^+e^-$ annihilation into hadrons $n_l=2$, while in
DIS or in the heavy flavor decays (\ref{seminclusive}) $n_l=1$.
Since hard collinear emissions are sensitive to the spin of the emitting
particles, the gluon function $B_g(\alpha)$ is not simply related to the
quark one $B(\alpha)$ \cite{demilio};
\item
the function $D(\alpha)$ represents emissions at large angle and small
energy from the heavy quark.
$D_1$ is the coefficient of that piece of the matrix element squared
which is singular in the small energy limit:
\begin{equation}
\label{drepr}
D_1 \, \alpha \, \frac{d\omega}{\omega}  \, d\theta^2 .
\end{equation}
The non logarithmic integration over the angle $\theta$
or the transverse momentum $k_t$
has been done and does not appear explicitly in eq.~(\ref{expresum});
the integration over the energy $\omega$ is rewritten as an integral over $z$.
$D_1=0$ in all the processes involving light partons only, as for
instance DIS, Drell-Yan (DY) or $e^+e^-$ annihilation into hadrons, while
it is not zero in all the processes containing at least one heavy quark,
such as for example the heavy flavor decays (\ref{seminclusive}).
Note that the effective coupling appearing in the $D$ terms is
$\alpha\left[Q^2(1-z)^2\right]$ and is therefore substantially larger
for $1-z\ll 1$ than the coupling entering the hard collinear terms, namely
$\alpha\left[Q^2(1-z)\right]$.

\end{itemize}
Eq.~(\ref{expresum}) is therefore a generalization of the $O(\alpha)$
result,
possessing a double logarithm coming from the overlap
of the soft and the collinear region and a single logarithm of soft or
collinear origin (see eqs.~(\ref{doublerepr}), (\ref{brepr}) and (\ref{drepr}))
\footnote{
Let us remember however that only two of the three functions
appearing in eq.~(\ref{expresum}) are independent \cite{cattren2}.}.
The functions $A(\alpha)$ and $B(\alpha)$ are believed to by universal,
i.e. process independent to any order in perturbation theory, as
they represent the development of a parton into a jet, i.e.
one-particle properties.
The function $D(\alpha)$ on the contrary is process-dependent, as it describes
soft emission at large angle, with interference contributions
from all the hard partons in the process, i.e. it describes
global properties of the hadronic final states.
Let us observe that $A_2$ and $D_2$, unlike $B_2$, do not have a $C_F^2$ contribution.
That is a consequence of the eikonal identity, which holds in the soft limit \cite{dok}.
According to this identity, the abelian contributions simply exponentiate
the lowest order $O(\alpha \, C_F)$ term, just like in QED.
That means that there are no higher order terms in the exponent $G_N$.
Because of similar reasonings, $A_3$ does not have a $C_F^3$ contribution.

Despite its supposed asymptotic nature,
the numerical values of the coefficients show a rather good convergence
of the perturbative series. Note that all the double-logarithmic coefficients
$A_i$ are positive, implying an increasing suppression with the order of
the expansion (up to the third one) of the rate in the
threshold region.
On the contrary, the single-logarithmic coefficients
$B_i$ and $D_i$  -- with the exception of $B_2$ --- are all negative
and therefore tend to enhance the rate in the threshold region \cite{lucavecchio}.
We have:
\begin{eqnarray}
A_1 &=& + \, 0.424413;
\\
A_2 &=& + \, 0.420947 \, - \, 0.0375264 \, n_f \, = \, 0.308367;
\\
A_3 &=& + \, 0.592067 \, - \, 0.0923137 n_f \, - \, 0.000398167 \, n_f^2
\, =  \, 0.311542;
\\
B_1&=& - \, 0.318310;
\\
B_2&=& + \, 0.229655 \, + \, 0.04020 \, n_f \, = \, 0.350269;
\\
D_1&=& - \, 0.424413;
\\
D_2&=& - \, 0.556416 \, + \, 0.002502 \, n_f \, = \, - \, 0.548911.
\end{eqnarray}
With our definition, the $\beta$-function coefficients are, as well known,
all positive.
\begin{eqnarray}
\beta_0&=& + \, 0.87535 \, - \, 0.05305 \, n_f \, = \, + \, 0.71620;
\\
\beta_1&=& + \, 0.64592 \, - \, 0.08021 \, n_f \, = \, + \, 0.40529;
\\
\beta_2&=& + \, 0.71986 \, - \, 0.140904 \, n_f \, + \, 0.003032 \, n_f^2
\, = \, + \, 0.324436.
\end{eqnarray}
In the last member we have assumed 3 active flavors ($n_f=3$).

Let us now discuss the computation of the coefficients entering the resummation formula.
The occurrence of a Sudakov form factor in semileptonic $B$ decays
was acknowledged originally in \cite{altetal},
where a simple exponentiation involving $A_1$ and $B_1+D_1$ was performed.
The coefficient $A_2$ was computed for the first time, as far as we know,
in \cite{A2}. It was denoted $A_1 \, K$ since it was considered
a kind of renormalization of the lowest-order contribution:
\begin{equation}
A_1 \, \alpha \, \rightarrow \,  A_1 \, \alpha (1\, + \, K \, \alpha).
\end{equation}
The coefficient $A_2$ was obtained from the soft-singular part of the
$q\rightarrow q$ two-loop splitting function \cite{twoloopkernels}, that is
as the coefficient of the $1/(1-z)$ term\footnote{This is exactly the same
procedure which has been followed to derive the third-order coefficient
$A_3$ from the three-loop splitting function \cite{A3}.}.
$A_2$ was subsequently recomputed in \cite{korrad} in the framework of Wilson line
theory, where the function $A(\alpha)$ has a geometrical meaning:
it is the anomalous dimension of a cusp operator, representing the radiation
emitted because of a sudden change of velocity of a heavy quark,
\begin{equation}
\Gamma_{cusp}(\alpha) \, = \, \sum_{n=1}^{\infty} \Gamma_{cusp}^{(n)}\,\alpha^n
\, = \, \Gamma_{cusp}^{(1)}\,\alpha \, + \,  \Gamma_{cusp}^{(2)}\,\alpha^2 \, + \, \cdots
\end{equation}
Indeed, it has been explicitly checked up to second order that these two functions
coincide:
\begin{equation}
\label{equality}
A(\alpha) \, = \, \Gamma_{cusp}(\alpha).
\end{equation}
Let us note that:
\begin{itemize}
\item
the theory of Wilson lines and Wilson loops;
\item
the eikonal or soft approximation in perturbative QCD;
\item
the heavy quark effective theory (HQET) and the large energy effective theory (LEET),
\end{itemize}
all involve basically the same structure, i.e. the same
propagators and vertices and the same amplitudes.
Since the same structure has been studied in different frameworks,
there is multiple notation and terminology for the same objects.
Let us stress however that in ordinary QCD the function $A(\alpha)$
is not an anomalous dimension, since it is not obtained from ultraviolet $1/\epsilon$
poles in renormalization constants but from infrared poles or from finite parts
of scattering amplitudes.
$A(\alpha)$ becomes an anomalous dimension {\it in} the effective theory
because the latter has additional ultraviolet divergencies with respect
to QCD. While in QCD one has to subtract only ultraviolet divergencies
related to coupling constant renormalization, in the effective theory one
has also to subtract {\it additional} ultraviolet divergencies related to the
cusp operators. A scheme dependence is therefore introduced in the effective
theory, which is not present in full QCD.
It seems to us therefore that the equality (\ref{equality}) is not guaranteed
{\it a priori} in higher orders and may require a specific scheme
for the subtractions in the effective theory.
At present, $A_3$ has only been derived in full QCD and not in the effective
theory.

The coefficient $B_2$ has been computed by means of the second order
correction to the inclusive DIS cross section, which contains the combination
$B_2+D_2^{DIS}$ (the DIS analogue of eq.~(\ref{refG21}), see later)
and by means of the third order correction,
which contains the different combination $B_2+2D_2^{DIS}$
(the DIS analogue of eq.~(\ref{refG32}), see later).
The knowledge of the fermionic contribution to the $O(\alpha^3)$ DIS
cross section was sufficient for a complete determination of $B_2$ \cite{B2},
with later checks offered by the complete computation \cite{A3,checknnlo}.

An incorrect value for the coefficient $D_2$ for heavy favor decays has been
obtained in the original computation in \cite{D2first},
where the technique to compute real and virtual diagrams in the
effective-theory in configuration space has also been developed.
In \cite{neub} the coefficient of the single logarithm in the radiative
decay (\ref{bsgamma}) to order $\alpha^2$ has been presented, from which the
correct value of $D_2$ can be extracted (let us note however that numerically
the two values are not very different).
A second order computation of heavy flavor fragmentation in ordinary QCD
was presented in \cite{fragmentation}, which allows the determination
of the sum $B_2+D_2^{frag}$ (the analogue of eq.~(\ref{refG21}), see later).
Using an identity relating the coefficient
for heavy flavor fragmentation with that one for heavy flavor decays,
and subtracting the known value for the
universal coefficient $B_2$, the correct value for $D_2$ was explicitly derived
in \cite{gardi0} (see also \cite{jafferandall}).
Still in \cite{gardi0}, by repeating the Wilson line computation of \cite{D2first},
errors were found and the same value of $D_2$ extracted from heavy flavor
fragmentation was re-obtained.
Recently, the second order contribution of the chromomagnetic operator
$O_7$ to the photon spectrum in the radiative decay
(\ref{bsgamma}) has been calculated \cite{O7twoloops}, confirming
these results (see also \cite{gardi}).

According to the previous remarks  concerning the relation between
$A(\alpha)$ and $\Gamma_{cusp}(\alpha)$, we believe
it is a non-trivial fact that the same value of $D_2$ is obtained with
two completely different methods:
\begin{itemize}
\item
a direct computation in the effective theory, which describes the soft region
only;
\item
an extraction from an ordinary QCD computation, which gives the sum
of the soft and the collinear contributions $B_2 + D_2$, by subtracting the collinear
contribution $B_2$ obtained from second order and third order DIS computations.
\end{itemize}

The following expansion holds true for the exponent:
\begin{equation}
\label{GNexpansion}
G_N(\alpha) \,=\, \sum_{n=1}^{\infty} \sum_{k=1}^{n+1} G_{n k}\alpha^n l^k
\, + \, O\left( \frac{1}{N} \right),
\end{equation}
where
\begin{equation}
l \, = \, \log N.
\end{equation}
The expansion of the logarithm of the form factor has a similar
structure in physical space and in $N$-space; roughly speaking,
going to $N$-space, $\log 1/u \, \rightarrow \, \log N$.

As already discussed, we are interested in the large-$N$ limit;
the $O(1/N)$ terms can be neglected in our leading-twist analysis.
A resummation of all the logarithmically-enhanced terms in (\ref{GNexpansion})
is at present unfeasible in QCD even in $N$-space,
so one has to rely on approximate schemes.
Let us discuss the fixed-logarithmic accuracy scheme:
\begin{itemize}
\item
Leading order (LO). One keeps in the exponent $G_N(\alpha)$ only the
leading power of the logarithm for each power of $\alpha$, i.e.
$k=n+1$:
\begin{equation}
G_N^{LO} \,=\, \sum_{n=1}^{\infty} G_{n\,n+1} \alpha^n l^{n+1} \,=\,
G_{12} \, \alpha \, l^2 \, + \, G_{23} \, \alpha^2 \, l^3 \, + \, O(\alpha^3).
\end{equation}
The coefficient function is kept in lowest order, i.e. $C^{LO}=1$
and the remainder function is completely neglected, i.e. $d^{LO}=0$;
\item
Next-to-leading order (NLO). One keeps in $G_N(\alpha)$ also the terms
with $n=k$, i.e.:
\begin{eqnarray}
G_N^{NLO} &=& \sum_{n=1}^{\infty} \left[ G_{n\,n+1} \, \alpha^n \, l^{n+1} \, + \,
G_{n\,n} \, \alpha^n l^n \right]
\nonumber\\
&=& G_{12} \alpha l^2 +  G_{11} \alpha l + G_{23} \alpha^2 l^3 +
G_{22} \alpha^2 l^2  + O(\alpha^3).
\end{eqnarray}
To $O(\alpha)$ one retains both the double and the single
logarithm. In general for each order in $\alpha$ one keeps the
principal two logarithms. One also keeps the $O(\alpha)$ terms
both in the coefficient function and in the remainder function:
\begin{equation}
C^{NLO} \, = \, 1 \, + \, \alpha \, C^{(1)}; ~~~~~~~~
d^{NLO} \, = \, \alpha \, d^{(1)}.
\end{equation}
The one-loop coefficient function is needed because of the
factorized form of the QCD form factor.
One has indeed a resummed expression of the form:
\begin{equation}
\left[ 1 \, + \, \alpha \, C^{(1)} \right] \, e^{ G_{12} \, \alpha l^2 \, + \,
\cdots}
\end{equation}
By expanding the exponent in powers of $\alpha$, a term coupling
the coefficient function and the double logarithm is obtained:
\begin{equation}
\alpha^2 \, C^{(1)} \, G_{12} \, l^2,
\end{equation}
which must be included in the NLO approximation;
\item
Next-to-next-to-leading order (NNLO). One keeps in $G_N$ also the
terms with $n=k-1$, i.e.:
\begin{eqnarray}
G^{NNLO}_N &=& \sum_{n=1}^{\infty} \left[ G_{n\,n+1} \alpha^n l^{n+1} +
G_{n\,n} \alpha^n l^n + G_{n\,n-1} \alpha^n l^{n-1} \right]
\nonumber\\
&=& G_{12} \alpha l^2 +  G_{11} \alpha l + G_{23} \alpha^2 l^3 +
G_{22} \alpha^2 l^2 + G_{21} \alpha^2 l  + G_{34} \alpha^3 l^4 +
G_{33} \alpha^3 l^3 + G_{32} \alpha^3 l^2 +
\nonumber\\
&+& O(\alpha^4).
\end{eqnarray}
To $O(\alpha^2)$, all the infrared logarithms
are included. In general, for each order in $\alpha$, one keeps
the principal three logarithms. The first omitted term is the single
logarithm to order $\alpha^3$.
One has also to keep the
$O(\alpha^2)$ terms both in the coefficient function and in the
remainder function:
\begin{equation}
C^{NNLO} \, = \, 1 \, + \, \alpha \, C^{(1)} \, + \, \alpha^2 \,
C^{(2)}; ~~~~~~~~
d^{NNLO} \, = \, \alpha \, d^{(1)} \, + \,\alpha^2 \, d^{(2)}.
\end{equation}
\end{itemize}
The classes of logarithms discussed above can be explicitly resummed by means
of a function series expansion of $G_N(\alpha)$ \cite{cattren1}:
\begin{eqnarray}
G_N(\alpha) \,=\, l \, g_1(\lambda) \, + \, \sum_{n=0}^{\infty} \alpha^n \, g_{2+n}(\lambda)
\, = \, \, l\, g_1(\lambda) + g_2(\lambda) + \alpha \, g_3(\lambda)  + \alpha^2 \, g_4(\lambda)+ \cdots,
\end{eqnarray}
where
\begin{equation}
\lambda\, = \, \beta_0 \, \alpha \, l.
\end{equation}
The $g_i(\lambda)$ are homogeneous functions of $\lambda$ and have
a series expansion around $\lambda = 0$:
\begin{equation}
g_i(\lambda) \, = \, \sum_{n=1}^{\infty} g_{i \,n}\, \lambda^n.
\end{equation}
In LO one needs only the function $g_1$, in NLO one need also $g_2$,
in NNLO also $g_3$ is needed and so on.
The explicit expressions read:
\begin{eqnarray}
g_1(\lambda) &=& -\frac{{A_1}}{2\,\beta_0\, \lambda}
    \left[ \left( 1 - 2\,\lambda \right) \,
       \log (1 - 2\,\lambda)
     - 2\,  \left( 1 - \lambda \right)
           \,\log (1 - \lambda)  \right];
\\
g_2(\lambda)&=& \frac{{D_1}}{2\,{{\beta}_0}} \log (1 - 2\,\lambda)
+ \frac{B_1}
   {{\beta}_0} \,\log (1 - \lambda )
+\frac{{A_2}}{ 2\,
     {{\beta}_0}^2} \,
     \left[ \log (1 - 2\,\lambda ) -
       2\,\log (1 - \lambda ) \right]  +
\\ \nonumber
&-&\frac{{A_1}\, {{\beta}_1} }{4\,{{{\beta}_0}}^3}\,
     \left[ 2\,\log (1 - 2\,\lambda ) +
       {\log^2 (1 - 2\,\lambda )} -
       4\,\log (1 - \lambda ) -
       2\,{\log^2 (1 - \lambda )} \right] \,
      +
\\ \nonumber
&+& \frac{{A_1}\,{{\gamma }_E}  }
     {{{\beta}_0}}\,
     \left[ \log (1 - 2\,\lambda ) -
       \log (1 - \lambda ) \right] +
\frac{A_1}{2 \, \beta_0} \left[\log(1 - 2\,\lambda )-2 \,\log (1 -
\lambda ) \right]\, \log \frac{\mu^2}{Q^2}.
\end{eqnarray}
The function $g_1(\lambda)$ in \cite{me} is in agreement with that one
obtained originally in \cite{akhouri}. $g_2(\lambda)$ in \cite{me}
differs instead from the corresponding $g_2^{sl}(\lambda)$
obtained in \cite{akhouri} and it is equal to the corresponding function
entering the $B\rightarrow X_s\gamma$ spectrum; the formalism we use makes
explicit the universality of soft gluon dynamics in semileptonic
and radiative decays.
The NNLO function $g_3$ has the rather lengthy expression:
\begin{eqnarray}
g_3(\lambda) &=&   -
  \frac{{D_2}\,\lambda }
   {{\beta_0}\,
     \left( 1 - 2\,\lambda  \right) } -
  \frac{2\,{D_1}\,{\gamma_E}\,
     \lambda }{1 - 2\,\lambda } + \frac{{D_1}\,{\beta_1}}{2\,
     {{\beta_0}}^2}\,
     \left( \frac{2\,\lambda }{1 - 2\,\lambda } +
       \frac{\log (1 - 2\,\lambda )}
        {1 - 2\,\lambda } \right)-\frac{{B_2}\,\lambda }
   {{\beta_0}\,
     \left( 1 - \lambda  \right) } -
  \frac{{B_1}\,{\gamma_E}\,
     \lambda }{1 - \lambda }
+\nonumber\\
&+&
  \frac{{B_1}}{{{\beta_0}}^2} \,{\beta_1}\,
     \left( \frac{\lambda }{1 - \lambda } +
       \frac{\log (1 - \lambda )}{1 - \lambda }
       \right)
%+\nonumber\\
%&-&
 - \frac{{A_3}}
     {2\,{{\beta_0}}^2}\,
     \left( \frac{\lambda }{1 - 2\,\lambda } -
       \frac{\lambda }{1 - \lambda } \right)
 -  \frac{A_2\,
       {\gamma_E} }{
       {\beta_0}}\,
       \left( \frac{1}{1 - 2\,\lambda } -
         \frac{1}{1 - \lambda } \right)
+\nonumber\\
&+&
  \frac{{A_2}\,{\beta_1}}{2\,{{\beta_0}}^3}\,
     \left( \frac{3\,\lambda }{1 - 2\,\lambda } -
       \frac{3\,\lambda }{1 - \lambda } +
       \frac{\log (1 - 2\,\lambda )}
        {1 - 2\,\lambda } -
       \frac{2\,\log (1 - \lambda )}{1 - \lambda }
       \right)
+\nonumber\\
&-&
  \frac{{A_1}\,{{\gamma_E}}^2 }{2}\,
     \left( \frac{4\,\lambda }{1 - 2\,\lambda } -
       \frac{\lambda }{1 - \lambda } \right)
- \frac{{A_1}\,{\pi }^2 }{12}\,
     \left( \frac{4\,\lambda }{1 - 2\,\lambda } -
       \frac{\lambda }{1 - \lambda } \right)
+\nonumber\\
 &-&
  \frac{{A_1}\,{\beta_2}}{4\,
     {{\beta_0}}^3}\,
     \left( \frac{2\,\lambda }{1 - 2\,\lambda } -
       \frac{2\,\lambda }{1 - \lambda } +
       2\,\log (1 - 2\,\lambda ) -
       4\,\log (1 - \lambda ) \right)
+\nonumber\\
&+& \frac{{A_1}\,{\beta_1}\,
     {\gamma_E} }{{{\beta_0}}^2}\,
     \left( \frac{1}{1 - 2\,\lambda } -
       \frac{1}{1 - \lambda } +
       \frac{\log (1 - 2\,\lambda )}
        {1 - 2\,\lambda } -
       \frac{\log (1 - \lambda )}{1 - \lambda }
       \right)
+\nonumber\\
&-&
  \frac{A_1\,{\beta_1}^2}{2\,{{\beta_0}}^4}\,
     \left( \frac{\lambda }{1 - 2\,\lambda } -
       \frac{\lambda }{1 - \lambda } -
       \log (1 - 2\,\lambda ) +
       \frac{\log (1 - 2\,\lambda )}
        {1 - 2\,\lambda }  \right. \nonumber \\
         &+& \left.
       \frac{{\log (1 - 2\,\lambda )}^2}
        {2\,\left( 1 - 2\,\lambda  \right) } +
       2\,\log (1 - \lambda ) -
       \frac{2\,\log (1 - \lambda )}
        {1 - \lambda } -
       \frac{{\log (1 - \lambda )}^2}{1 - \lambda }
       \right)
+\nonumber\\
&-&
  \frac{{D_1}\,\lambda }{{\beta_0}\,\left( 1 - 2\,\lambda  \right) } \,
     \log \frac{{\mu }^2}{Q^2} -
  \frac{{B_1}\,\lambda }{{\beta_0}\,\left( 1 - \lambda  \right) }\,
     \log \frac{{\mu }^2}{Q^2}
%+\nonumber\\
%&-&
 - \frac{{A_2}}{{{\beta_0}}^2}\,
     \left( \frac{\lambda }{1 - 2\,\lambda } -
       \frac{\lambda }{1 - \lambda } \right) \,
     \log \frac{{\mu }^2}{Q^2}
+\nonumber\\
&-& \frac{{A_1}\,
     {\gamma_E} }{{\beta_0}}\,
     \left( \frac{2\,\lambda }{1 - 2\,\lambda } -
       \frac{\lambda }{1 - \lambda } \right) \,
     \log \frac{{\mu }^2}{Q^2}
+\nonumber\\
&+& \frac{{A_1}\,
     {\beta_1} }{{{\beta_0}}^3}\,
     \left( \frac{\lambda }{1 - 2\,\lambda } -
       \frac{\lambda }{1 - \lambda } +
       \frac{\log (1 - 2\,\lambda )}{2} +
       \frac{{\lambda }\,
          \log (1 - 2\,\lambda )}{1 - 2\,\lambda }
        - \log (1 - \lambda )
- \frac{\lambda \,\log (1 - \lambda )}
        {1 - \lambda } \right) \,
     \log \frac{{\mu }^2}{Q^2}\!
+\nonumber\\
&-& \frac{{A_1}}{2\,
     {\beta_0}}\,
     \left( \frac{2\,{\lambda }^2}
        {1 - 2\,\lambda } -
       \frac{{\lambda }^2}{1 - \lambda } \right) \,
     {\log^2 \frac{{\mu }^2}{Q^2}}.
\end{eqnarray}
The function $g_3(\lambda)$ was originally computed in
\cite{ugnnlo}, where the first NNLO resummation in  heavy flavor
decays was presented.
At the time of that work, not all the fixed-order
computations were available from which to extract the
coefficients entering the resummation formula, namely
$A_3$, $B_2$ and $D_2$.
A numerical estimate of the three-loop coefficient $A_3$
was used, which was
obtained in \cite{vanervogt} by fitting the known moments of the
3-loop splitting kernels and which has been later confirmed by the
exact analytic evaluation \cite{A3}.
As far as $B_2$ is concerned, an approximation based on the
$q \rightarrow q$ splitting function at two loops has been assumed,
which was shown to be rather poor
by the subsequent exact computation in \cite{B2}.
The coefficient $D_2$ was taken from its original computation in
\cite{D2first}.
There is a misprint in $g_3(\lambda)$ in \cite{ugnnlo}
in two terms proportional to $A_1 \beta_2$:
$\log[1-\lambda]-1/2\log[1-2 \lambda]$ has to be multiplied by a factor 2,
as found indeed in the recent recomputation of the $\mu$-independent
terms \cite{gardi}.
With the misprint, the terms proportional to $A_1 \beta_2$ would indeed appear
at $\alpha^3$, while they have to appear only at order $\alpha^4$,
as shown correctly in the $\alpha$ expansion of the $g_3$ in eq.~(42)
of \cite{ugnnlo}.

Let us note that the soft terms, i.e. the terms proportional to the coefficients
$A_i$ and $D_i$, have the singularity closest to the origin in $\lambda=1/2$
while the collinear terms, proportional to $B_i$, have only a singularity in
$\lambda=1$.

\subsection{Inverse transform to physical space}

The original form factor in $u$ space is recovered by an
inverse Mellin transform:
\begin{equation}
\sigma(u;\,\alpha) \, = \, \int_{c-i\infty}^{c+i\infty}
\frac{dN}{2\pi i} (1-u)^{-N} \, \sigma_N(\alpha),
\end{equation}
where $c$ is a real constant chosen in such a way that all the
singularities of $\sigma_N$ lie to the left of the integration contour.
The inverse transform can be done to any given logarithmic accuracy
in closed analytic form, where now the logarithmic accuracy is defined
as before but in terms of powers of $\alpha$ and $L=\log 1/u$ instead
of $l=\log N$.
To NNLO accuracy, one can write \cite{ugnnlo}
\footnote{
A factor $1-u\approx 1$ has been neglected in our leading twist accuracy.
}:
\begin{equation}
\label{SigmaNNLO}
\Sigma \left[u;\alpha\right] \,=\,
\frac{  e^{ L \, g_1(\tau) \, +  \, g_2(\tau)  }  }
{  \Gamma\left[1 - h_1(\tau) \right]  } \delta\Sigma,
\end{equation}
where
\begin{equation}
\tau \, = \, \beta_0 \, \alpha \, L
\end{equation}
and we have defined
\begin{equation}
h_1(\tau) \, = \, \frac{d}{d\tau}\left[\tau g_1(\tau)\right]
\, = \, g_1(\tau) + \tau \, g_1'(\tau).
\end{equation}
$\delta\Sigma$ is a NNLO correction factor which can be set
equal to one in NLO:
\begin{equation}
\delta\Sigma_{NLO} \, = \, 1.
\end{equation}
Its NNLO expression reads:
\begin{equation}
\delta\Sigma \,=\, S / S|_{L\rightarrow 0}
\end{equation}
with
\begin{equation}
S \,=\, e^{ \alpha \, g_3(\tau) }
\Bigg\{
1 \, + \, \beta_0 \, \alpha \, g_2'(\tau) \,
\psi\left[1 - h_1(\tau) \right]
\, + \, \frac{1}{2} \beta_0 \, \alpha \, h_1'(\tau)
\Big\{
\psi^2\left[1-h_1(\tau)\right]
- \psi'\left[1-h_1(\tau)\right]
\Big\}
\Bigg\}.
\end{equation}
In \cite{ugnnlo} inhomogeneous terms were included in
$\delta\Sigma$, which have been subtracted here.
$\Gamma(x)$ is the Euler Gamma function and
\begin{equation}
\psi(x) \,=\, \frac{d}{d x} \log \Gamma(x)
\end{equation}
is the digamma function.

\noindent
Expanding the r.h.s. of eq.~(\ref{SigmaNNLO}) up to third order, one obtains the
following relations:
\begin{eqnarray}
G_{12} &=& -\frac{1}{2} A_1;
\\
G_{11} &=& - ( B_1 + D_1 );
\\
G_{23} &=& - \frac{1}{2} A_1 \beta_0;
\\
G_{22} &=& - \frac{1}{2} A_2 - \frac{1}{2} \beta_0 (B_1 + 2\,D_1) - \frac{1}{2} A_1^2 z(2);
\\
\label{refG21}
G_{21} &=& - (B_2 + D_2) - A_1 \left( B_1 + D_1 \right) z(2) - A_1^2 z(3);
\\
G_{34}&=& -\frac{7}{12} A_1 \beta_0^2;
\\
G_{33}&=& - A_2 \beta_0 - \frac{1}{2}  A_1 \beta_1 - \frac{1}{3}\beta_0^2
\left(  B_1
+ 4\,  D_1 \right) - \frac{3}{2}  A_1^2 \beta_0 z(2) +  \frac{1}{3}  A_1^3 z(3);
\\
\label{refG32}
G_{32}&=&
- \frac{1}{2} A_3
%-
%  \frac{{\beta_1}\,
%     {B_1}}{2}
-
  {\beta_0}\, ({B_2} + 2\, D_2)
-\frac{\beta_1}{2}\,
  (B_1+2\, {D_1})
%-
%  2\,{\beta_0}\,
%   {D_2}
-
{A_1}\,{A_2}\,z(2)
- \frac{{A_1}\,
     {\beta_0}}2\,
     (5\,{B_1}\,
      + 7\,
     {D_1})\,z(2)
 +
\nonumber\\
&+& \frac{{{A_1
        }}^3}{4}\,z(4) -
  \frac{9\,{{A_1}}^2\,
     {\beta_0}\,
     {z}(3)}{2} +
  {{A_1}}^2\,
   ({B_1}\,+D_1) z(3)
%   {z}(3) +
%  {A_1}^2\,
%   {D_1}\,
%   {z}(3)
,
\end{eqnarray}
where $z(a)=\sum_{n=1}^{\infty} 1/n^a$ is Riemann Zeta function
with $z(2)=\pi^2/6=1.64493\cdots$, $z(3)=1.20206\cdots$ and $z(4)=\pi^4/90=1.08232\cdots$.
Note that the leading coefficients $G_{23}$ and $G_{34}$ involve products of the one-loop
coefficients $A_1$ and $\beta_0$ only.
The explicit expressions of the $G_{ij}$ read:
\begin{eqnarray}
G_{12} &=& - \frac{C_F}{2\pi};
\\
G_{11} &=& \frac{7 C_F}{4\pi};
\\
G_{23} &=&
- \frac{ C_F }{8\,{\pi }^2} \,
      \left( \frac{11\,C_A}{3} -
        \frac{2\,n_f}{3} \right)  ;
\\
G_{22} &=&
\frac{C_F }{4\,{\pi }^2}\,
    \left[  C_A\, \left( \frac{95}{72} +
         z(2)%\frac{{\pi }^2}{6}
\right)
- \frac{13\,n_f}{36} -
2\, C_F\, z(2) %     \frac{C_F\,{\pi }^2}{3}
\right] ;
\\
G_{21} &=&
\frac{C_F}{6\,\pi^2}\,
    \left[ n_f\, \left(
       - \frac{85}{24} +
        z(2) %\frac{{\pi }^2}{6}
\right)  +
      C_A\,
       \left( \frac{905}{48} -
       \frac{17}{2}\,z(2) % \frac{17\,{\pi }^2}{12}
 -
         \frac{3\,z(3)}{2} \right)
       + C_F\,
       \left( \frac{9}{16} + 6\,z(2) + %{\pi }^2 +
         3\,z(3) \right)  \right];
\\
G_{34} &=&
\frac{C_F}{48\,{\pi }^3}\,
    \left( - \frac{847\,
         {C_A}^2}{36} +
      \frac{77\,C_A\,
         n_f}{9} -
      \frac{7\,{n_f}^2}
       {9} \right);
\\
G_{33} &=&
\frac{ C_F }{4\,{\pi }^3}\,
    \Bigg[
      - \, \frac{{n_f}^2}
       {108} + C_A\,
       n_f\,
       \left( \frac{20}{27} -
         \frac{z(2)}{3} \right)  +
      {C_A}^2\,
       \left( - \frac{1261}
            {432}  +
         \frac{11\,z(2)}{6} \right) \,+\,
\nonumber\\
&-& \frac{11\,
            z(2)}{2} C_A\, C_F\,+
          C_F \, n_f\,
          \left( \frac{1}{4} +
            z(2) \right)  +
      \frac{4\,{C_F}^2\,
         z(3)}{3}
\Bigg];
\\
G_{32} &=&
\frac{C_F}{4\,{\pi }^3}\,
\Bigg[
{n_f}^2\,
       \left( \frac{275}{648} -
         \frac{z(2)}{9} \right)  +
      C_A\,n_f\,
       \left( - \frac{5399}
            {1296}  +
         \frac{4\,z(2)}{3} -
         \frac{z(3)}{6} \right)  \, + \,
\nonumber\\
&+& {C_A}^2\,
       \left( \frac{21893}{2592} -
         \frac{119\,z(2)}{36} +
         \frac{77\,z(3)}{12} -
         \frac{11\,z(4)}{4} \right)  +
       {C_F}^2\,
       \Big( -7\,z(3) + z(4) \Big) \, + \,
\nonumber\\
&+& C_F\, n_f\,
          \left( \frac{19}{48} -
            \frac{71\,z(2)}{36} + z(3)
            \right)  +
         C_A\, C_F\,
          \left( \frac{11}{32} +
            \frac{685\,z(2)}{72} -
            11\,z(3) + 5\,z(4) \right)
\Bigg].
\end{eqnarray}
The numerical values of the coefficients show a good convergence
of the perturbative series also in configuration space:
\begin{eqnarray}
G_{12} & = & - \, 0.212207 ;
\\
G_{11} & = & + \, 0.742723 ;
\\
G_{23} & = & - \, 0.185756 \, + \, 0.011258\, n_f \, = \, - \, 0.151982;
\\
G_{22} & = & + \, 0.152206 \, - \, 0.012196\, n_f \, = \, + \, 0.115618;
\\
G_{21} & = & + \, 0.628757 \, - \, 0.0427065\, n_f \, =\, + \, 0.500638;
\\
G_{34}& = & - \, 0.189702 \, + \, 0.022994\, n_f \, - \, 0.0006968\, n_f^2 \, = \, - \, 0.126990;
\\
G_{33}& = & - \, 0.349055 \, + \, 0.033368\, n_f \, - \, 0.0000995\, n_f^2 \, = \, - \, 0.249846;
\\
G_{32}& = & + \,0.96117 \, - \, 0.09368\, n_f \, + \, 0.0025974\, n_f^2 \, = \, + \, 0.703506,
\end{eqnarray}
where on the last member of the r.h.s. we have set $n_f=3$.

\section{Distribution in the hadronic variables}
\label{hadrvars}
The distribution in the hadronic variables $u$ and $w$ is obtained integrating the
triple differential distribution (\ref{triplefin})
over the electron energy $\overline{x}=1-x$. The integration range is
\begin{equation}
\label{exrange}
\overline{x}_1(w,u) \,\le\, \overline{x}  \,\le\, \overline{x}_2(w,u),
\end{equation}
where:
\begin{equation}
\overline{x}_1(w,u)\,=\,\frac{w\,u}{1+u} ~~~~~~~~{\rm and }~~~~~~~~
\overline{x}_2(w,u)\,=\,\frac{w}{1+u}.
\end{equation}
Let us use the second method of integration of the triple-differential distribution
discussed at the end of sec.~(\ref{triplodiff}), i.e. let us neglect at first the
remainder function.
Since the QCD form factor $\sigma\left[u;\,\alpha(w\,m_b)\right]$ does not depend
on the electron energy $\overline{x}$, the integration only involves the coefficient
function:
\begin{equation}
\int_{\overline{x}_1}^{\overline{x}_2} d\overline{x} \, C(\overline{x},w;\,\alpha).
\end{equation}
We eliminate small terms $O(u)$ from the integral above by integrating over the
range which is the limit $u\rightarrow 0$ of (\ref{exrange}):
\footnote{The latter
are actually the integration regions for $e^+e^-\rightarrow q +
\overline{q}+g$ with massless quarks.}
\begin{equation}
\overline{x}_1(w,0) \, \le \, \overline{x} \, \le \, \overline{x}_2(w,0).
\end{equation}
In fact, these terms $O(u)$, when multiplied with the plus distributions of $u$ contained
in the QCD form factor $\sigma(u;\,\alpha)$, give at worse terms of the form $\log u$,
which miss the $1/u$ enhancement and therefore are to be considered as ``small''.
Let us define therefore the coefficient function of the double hadronic
distribution as:
\begin{equation}
C_H\left(w;\,\alpha\right)
\, = \, \int_0^w d \overline{x} \, C\left(w,\overline{x};\alpha\right),
\end{equation}
having the usual $\alpha$ expansion
\begin{equation}
C_H\left(w;\,\alpha\right) \, = \, C^{(0)}_H (w) \, + \, \alpha \, C^{(1)}_H(w)
\, + \, \alpha^2 \, C^{(2)}_H(w) \, + \, O(\alpha^3).
\end{equation}
One easily obtains:
\begin{eqnarray}
C_H^{(0)}(w) &=&  2 w^2 (3-2w);
\\
C_H^{(1)}(w) &=& \frac{C_F}{\pi} w^2 \left\{
- (9-4w) \log w +
2 (3-2w) \left[ {\rm Li}_2(w) + \log w \log(1-w) - \frac{35}{8}  \right]
\right\}.
\nonumber
\end{eqnarray}
The first two orders of the coefficient function vanish as $w^2$
for $w\rightarrow 0$, implying a suppression of the states with
a small hadronic energy (i.e. with a small hard scale), as
anticipated in the introduction.

The resummed distribution in the hadronic variables $u$ and $w$ then reads:
\begin{equation}
\label{hadrvar}
\frac{1}{\Gamma}\frac{d^2\Gamma}{du dw} \, = \, C_H\left[w;\,\alpha(m_b)\right] \,
\sigma\left[u;\,\alpha(w\,m_b)\right] \, + \, d_H\left[u,w;\,\alpha(m_b)\right],
\end{equation}
where the remainder function has an expansion analogous to the one in the triple
differential distribution:
\begin{equation}
d_H\left(u,w;\,\alpha \right) \, = \,
\alpha \, d_H^{(1)}(u,w) \, + \,
\alpha^2 \, d_H^{(2)}(u,w) \, + \,
O(\alpha^3).
\end{equation}
Expanding to first order the above distribution and comparing (matching) with
the known $O(\alpha)$ distribution, the following remainder function is
obtained --- an over-all factor $C_F/\pi$ is omitted:
\begin{eqnarray}
d_H^{(1)}(w,u) =
&-& \, \frac{4\,w^6\,\log \, u}{{\left( 1 + u \right) }^7}
+ \frac{4\,w^2\,\left( 3 - 2\,w \right) \,\log \, u}{1 + u}
- \frac{ 32\,w^5 - 10\,w^6\,\log \, u}{{\left( 1 + u \right) }^6} +
\nonumber\\
&+& \frac{ 3\,w^2\,\left( 14 - 6\,w - 5\,w^2 \right)
         - 2\,w^3\,\left( 3 - 4\,w \right) \,
      \log \, u}{{\left( 1 + u \right) }^2} +
\nonumber\\
&+& \frac{ 20\,w^3\,
      \left( 2 + w \right) \,
      \left( 1 - 2\,w \right) -
     w^4\,\left( 9 - 18\,w - 2\,w^2 \right) \,\log \, u}{
       {\left( 1 + u \right) }^4} +
\nonumber\\
&+&  \frac{64\,w^5 +
     2\,w^4\,\left( 3 - 6\,w -
        4\,w^2 \right) \,\log \, u}{
       {\left( 1 + u \right) }^5} +
\nonumber\\
&-& \frac{ 4\,w^3\,\left( 10 - 15\,w -
        2\,w^2 \right)  - w^3\,\left( 12 - 13\,w - 6\,w^2 \right) \,\log \, u}
     {{\left( 1 + u \right)}^3}.
\end{eqnarray}
Eq.~(\ref{hadrvar}) provides a complete NLO resummation of the distribution in the
two hadronic variables $u$ and $w$, from which the distribution in any other pair of
hadronic variables can be obtained by a change of variables.
One can insert in eq.~(\ref{hadrvar}) the NNLO form factor $\sigma$, whose
properties have been discussed in sec.~(\ref{secthresum}), allowing
an approximate NNLO resummation.
In fact, for a complete NNLO resummation, one also needs the second order
corrections to the coefficient function $C_H^{(2)}(w)$ and the remainder function
$d_H^{(2)}(u,w)$, which are unknown at present.

\section{Hadron energy spectrum}
\label{hadspec}

The distribution in the total hadron energy $w$ is obtained
by integrating the distribution in the hadronic variables (\ref{hadrvar}).
The integration range in $u$ is:
\begin{equation}
\max(0,w-1)\,\le\, u \,\le\, 1.
\end{equation}
Since the coefficient function $C_H(w;\,\alpha)$ does not depend
on $u$, the integration only involves the QCD form factor and
the remainder function:
\begin{equation}
\label{nonfactori}
\frac{1}{\Gamma}\frac{d\Gamma}{dw} \,=\,
C_H\left(w;\,\alpha\right) \Big\{1\, - \, \theta(w-1)
\Sigma\left[w-1;\,\alpha(w\,m_b)\right]\Big\}
\, + \,
\int_{\max(0,w-1)}^1 dw \, d_H(u,w;\,\alpha),
\end{equation}
where $\Sigma(u;\,\alpha)$ is the partially-integrated form factor defined in
section (\ref{secthresum}).

Because of the $\theta(w-1)$ multiplying $\Sigma(w-1;\,\alpha)$, there are large
logarithms only for $w>1$, as anticipated in the qualitative
discussion in the introduction.
We may therefore consider the parts of the spectrum for $w<1$ and $w>1$
as two different spectra, merging in the point $w=1$.
Let us consider the simpler case $w<1$ first.
Since, as already noted, there are no large logarithms, no resummation is
required and the $O(\alpha)$ fixed-order result coincides with the NLO one.
There is no QCD form factor and therefore there is no way to distinguish between
the coefficient function and the remainder function.
The spectrum for $w<1$ can then be written as an ordinary $\alpha$
expansion:
\begin{equation}
\frac{1}{2\Gamma}\frac{d\Gamma}{dw} \, = \, L(w;\,\alpha)
~~~~~~~~~~~~~~~~~~~~~~~~~~~~~~~~~~~~~~~~(w<1),
\end{equation}
where:
\begin{equation}
L(w;\,\alpha) \, = \,
L^{(0)}(w) \, + \, \alpha \, L^{(1)}(w) \, + \, \alpha^2 \, L^{(2)}(w)
\, + \, O(\alpha^3).
\end{equation}
The first two orders read \cite{enspec,ndf}:
\begin{eqnarray}
L^{(0)}(w) &=& w^2(3-2w);
\\
L^{(1)}(w) &=& \frac{C_F}{\pi}
\Bigg\{
- w^2(3-2w)\left[\frac{25}{8} + {\rm Li}_2(1-w) \right] \, +
\nonumber\\
&& ~~~~~~~
+\,\frac{1}{720}w^2\left(4w^4-42w^3+585w^2-3720w+4860+1440w\log w-3240 \log w \right)
\Bigg\}.
\end{eqnarray}

Let us now consider the more interesting case $w>1$, where
resummation is effective and one has to keep the resummed form of
the distribution in (\ref{nonfactori}). In a minimal scheme we
have to subtract small terms from the first term on the r.h.s. of
eq.~(\ref{nonfactori}), since the form factor must contain large
logarithms only. This is done setting $w=1$ in the argument of the
coupling entering the form factor $\Sigma$ as well as in the
coefficient function $C_H$, obtaining the simpler expression:
\begin{equation}
\label{basicw}
\frac{1}{\Gamma}\frac{d\Gamma}{dw} \,=\,
C_H\left(1;\,\alpha\right) \Big\{1\, - \,
\Sigma\left[w-1;\,\alpha(m_b)\right]\Big\}
\, + \, \cdots,
\end{equation}
where the dots denote terms not containing large logs of $w-1$.
Let us prove the legitimacy of the transformation from (\ref{nonfactori})
to (\ref{basicw}).
As far as the argument of the coupling is concerned, we expand
the QCD form factor $\Sigma$ in powers of $\alpha(w\,m_b)$.
One obtains terms of the form
\begin{equation}
\alpha(w\,m_b) \, \log^2(w-1) \, = \, \alpha(m_b) \log^2(w-1)
\, - \, 2 \beta_0 \alpha(m_b)^2 \log w \log^2(w-1)
\, + \, 4 \beta_0^2 \alpha(m_b)^3 \log^2 w \log^2(w-1)
 \, + \, \cdots,
\end{equation}
where on the r.h.s. an expansion of $\alpha(w\,m_b)$ around the point $w=1$
has been performed.
All the terms on the r.h.s except the first one vanish for $w\rightarrow 1^+$,
therefore they are not large logarithms and can be dropped. The only large logarithm
is the first term on the r.h.s., which is obtained by setting $w=1$ in the coupling
in the original expression on the l.h.s.
All this implies that the coupling can be evaluated
in the infrared-singular point $w=1$.
As far as the coefficient function is concerned, one just notices that the neglected terms,
\begin{equation}
\Big[
C_H\left(w;\,\alpha\right) \, - \, C_H\left(1;\,\alpha\right)
\Big]
\Big\{1 \, - \,
\Sigma\left[w-1;\,\alpha(m_b)\right]\Big\},
\end{equation}
are again vanishing for $w\rightarrow 1^+$,
because $C_H\left(w;\,\alpha\right) - C_H\left(1;\,\alpha\right)=O(w-1)$
and therefore can be neglected in this limit.

In NLO one has also to add a remainder function to
be determined via a matching procedure. That, as already discussed in other cases,
is in order to take into account also the region $w-1 \sim O(1)$.
One then has the resummed expression:
\begin{equation}
\label{wg1}
\frac{1}{2\Gamma}\frac{d\Gamma}{dw} \,=\, C_W\left(\alpha\right) \Big\{1 \, - \,
\Sigma\left[w-1;\,\alpha(m_b)\right]\Big\}
\, + \, H(w;\,\alpha)
~~~~~~~~~~~~~~~~~~~(w>1),
\end{equation}
where we have defined:
\begin{equation}
\label{automatic}
C_W(\alpha) \, \equiv \, \frac{1}{2}\, C_H(1;\,\alpha_s).
\end{equation}
The coefficient function and the remainder function have a standard $\alpha$ expansion:
\begin{eqnarray}
C_W(\alpha) &=& 1 \, + \, \alpha \, C_W^{(1)} \, + \, + \, \alpha^2 \, C_W^{(2)} \,+\, O(\alpha^3),
\\
 H(w;\,\alpha) &=& \alpha \, H^{(1)}(w) \, + \, \alpha^2 \, H^{(2)}(w) \, + \, O(\alpha^3).
\end{eqnarray}
The first order correction to the coefficient function reads:
\begin{equation}
\label{CW1}
C_W^{(1)} \, = \, \frac{C_F}{\pi} \left(\frac{\pi^2}{6} \, - \, \frac{35}{8}\right)
\, = \,  - \, 1.15868.
\end{equation}
Note that $C_W^{(1)}$ is negative and has a rather large size; for $\alpha(m_b)=0.22$
it gives a negative correction of $~\approx -\, 25\%$.
By using the matching procedure described at the end of sec.~(\ref{secthresum}),
we obtain:
\begin{eqnarray}
H^{(1)}(w) & = & \frac{C_F}{\pi}
\Big\{
\,-\,\frac{1}{2}\left( 2 w + 1 \right) (w-1)^2 \log^2 (w-1)
- \frac{1}{3}(w-1)(2w^2-w-4)\log(w-1) \, +
\nonumber\\
&&~~~~~~~
-\,w^2(3-2w)\left[ 2{\rm Li}_2\left(1-w\right) +2\log(w-1) \log w + \frac{\pi^2}{6} \right] \, +
\nonumber\\
&&~~~~~~~
+\,\frac{1}{720}(2-w)(4w^5-34w^4+517w^3-2946w^2+3798w+1248)
\Big\}.
\end{eqnarray}
The above function is positive in all the kinematical range $1<w<2$ and
goes to zero for $w\rightarrow 2$, as expected on the basis of the vanishing
of the phase space in this point.

Let us make a few remarks about eq.~(\ref{wg1}).
If we expand the r.h.s. of eq.~(\ref{wg1}) in powers of $\alpha$, we find that
$C_W^{(1)}$ only appears in order $\alpha^2$ --- this occurs because
the form factor multiplying the coefficient function is in this case
$1-\Sigma = O(\alpha)$ and not $\Sigma = O(1)$.
At present, only a full $O(\alpha)$ computation is available, implying that
$C_W^{(1)}$ cannot be determined by the matching: only the
remainder function can be fixed by this procedure.
The value of $C_W^{(1)}$ came out ``automatically'' as a consequence
of our resummation formula (see. eq.~(\ref{automatic})). There is however another method
to fix $C_W^{(1)}$: we require that the resummed spectrum is continuous in $w=1$.
Since $\Sigma(w-1)\rightarrow 0$ for $w\rightarrow 1^+$, we obtain the equation:
\begin{equation}
1 \, + \, \alpha \, L^{(1)}(1) \, = \,
1 \, + \, \alpha \left[ C_W^{(1)} \, + \, H^{(1)}(1) \right],
\end{equation}
to be solved in $C_W^{(1)}$:
\begin{equation}
C_W^{(1)} \, = \, L^{(1)}(1) \, - \, H^{(1)}(1)
\end{equation}
and giving again the value (\ref{CW1}).
The condition of continuity of the resummed spectrum
in $w=1$ is very reasonable from the physical viewpoint and it is
remarkable that the two methods give the same value for the coefficient
function.

Even though we are considering a differential spectrum,
its resummation involves, as we have explicitly seen,
the partially integrated form factor.
$\Sigma$ usually enters event fractions in expressions of the form
\begin{equation}
\label{genresum}
R(y;\,\alpha) \, = \, C(\alpha)\,\Sigma(y;\,\alpha)\, + \,
D(y;\,\alpha),
\end{equation}
with a remainder function vanishing for $y \rightarrow 0$, where
$y$ is a general kinematical variable entering the large
logarithms $\log 1/y$. In the case of the hadron energy spectrum,
its resummation is different from (\ref{genresum}) because it
involves the combination $1-\Sigma$ instead of $\Sigma$: there is
an additive constant, namely one, which makes the spectrum non
vanishing for $w\rightarrow 1^+$, as it should. It seem however
reasonable to impose the vanishing of the remainder function
$H(w;\alpha)$ for $w\rightarrow 1^+$ also in this case. The
previous factorization scheme does not satisfy this condition,
because:
\begin{equation}
H^{(1)}(1) \, = \, \frac{C_F}{\pi} \, \left(\frac{2587}{720}
-\frac{\pi^2}{6} \right).
\end{equation}
We can construct an improved scheme satisfying this condition
by introducing two coefficient functions instead of one:
\begin{equation}
\label{wg1impr}
\frac{1}{2\Gamma}\frac{d\Gamma}{dw} \,=\, C_{W1}\left(\alpha\right) \Big\{1 \, - \,
C_{W2} \left(\alpha\right)\,
\Sigma\left[w-1;\,\alpha(m_b)\right]
\, + \, \tilde{H}(w;\,\alpha)
\Big\}
~~~~~~~~~~(\mathrm{improved~scheme},~w>1),
\end{equation}
where the new remainder function, vanishing in $w=1$, reads:
\begin{equation}
\tilde{H}(w;\,\alpha)\,=\, H(w;\,\alpha)\,-\,H(1;\,\alpha).
\end{equation}
The coefficient functions have the usual fixed-order expansions:
\begin{eqnarray}
C_{W1}\left(\alpha\right) &=& 1 \, + \, \alpha \, C_{W1}^{(1)}
\, + \, \alpha^2 \, C_{W1}^{(2)} \, + \, O(\alpha^3);
\\
C_{W2}\left(\alpha\right) &=& 1 \, + \, \alpha \, C_{W2}^{(1)}
\, + \, \alpha^2 \, C_{W2}^{(2)} \, + \, O(\alpha^3).
\end{eqnarray}
By imposing the continuity in $w=1$ as in the previous scheme,
we obtain for the first coefficient function at first order in $\alpha$:
\begin{equation}
C_{W1}^{(1)} \, = \, L^{(1)}(1) \,=\, -\, \frac{C_F}{\pi} \, \frac{563}{720}
\,= \, -\,0.331868.
\end{equation}
The second coefficient function is obtained by imposing the usual matching
with the first order computation:
\begin{equation}
C_{W2}^{(1)} \, = -\, H^{(1)}(1) \,=\,-\, \frac{C_F}{\pi} \, \left(\frac{2587}{720}
-\frac{\pi^2}{6} \right)
\,= \, -\,0.826808.
\end{equation}
The improved resummed expression (\ref{wg1impr}) is positive in all the kinematical range $1<w<2$
and vanishes for $w\rightarrow 2$.

We can compare the hadron energy spectrum for $w>1$ given in eq.~(\ref{wg1}) or in
eq.~(\ref{wg1impr})  with the
hadron mass distribution in the radiative decay (\ref{bsgamma})
given in eq.~(\ref{radsum2}).
The hadron energy distribution contains $\Sigma$, i.e.
just the integral of the form factor $\sigma$ entering the radiative
decay spectrum. The hadron energy spectrum is therefore a very good
quantity on the theoretical side --- it is exceptional in this
respect --- being directly connected, via integration, to the radiative decay.
By that we mean that the connection between the two spectra only
involves short-distance coefficients. As show in \cite{wip},
this is to be contrasted with the case of other single-differential
spectra.

%
%   VALORE MEDIO DI w
%

\subsection{Average energy}

As discussed in the introduction,
the infrared singularity in $w=1$ of the $O(\alpha)$ spectrum is
integrable, so one can calculate directly the average hadronic energy as a truncated
expansion in $\alpha$:
\begin{equation}
\langle w \rangle \, = \, \frac{7}{10}
\left[
1 \, + \, \frac{\alpha \, C_F}{\pi} \, \frac{137}{840} \right]
\, = \, 0.71.
\end{equation}
The $O(\alpha)$ correction is very small, of the order of $1\%$, due to a large cancellation between
the contribution for $w<1$, which is negative, and the one for $w>1$, which is positive.
Setting for instance $m_b = m_B$ one obtains in leading order:
\begin{equation}
\langle E_X \rangle \, = \, \frac{7}{10} \, \frac{m_B}{2} \, = \, 1.843\, {\rm GeV}
\end{equation}
with a tiny first-order correction of $+\,26$ MeV.
This quantity can be directly compared with the experimental value.
In the radiative decay (\ref{bsgamma}) there is
a larger final hadronic energy: in lowest order
\begin{equation}
\langle E_X \rangle_{B\rightarrow X_s \gamma} \, = \, \frac{m_B}{2} \, = \, 2.634\, {\rm GeV.}
\end{equation}
The average hadronic energy is $\sim 30 \%$ larger in the radiative decay
than in the semileptonic decay,
in line with the qualitative discussion about the differences of the two decays
given in the introduction.

%
%   TAGLIO SU m_X
%

\subsection{Upper cut on hadron masses}

In experimental analysis an upper cut on invariant masses
\begin{equation}
m_X  \, < \, \overline{m}_{X}
\end{equation}
is imposed in order to kill the large background from semileptonic $b\rightarrow c$
transitions. Let us define:
\begin{equation}
k\, = \, 2 \frac{\overline{m}_{X}}{m_b}.
\end{equation}
In practice, $\overline{m}_{X}=1.6\div 1.8$ GeV, so we can assume $k<1$.
A leading order evaluation of the spectrum with the above cut gives:
\begin{equation}
\frac{1}{2\Gamma}\frac{d\Gamma}{dw} \, = \, \Bigg\{
\begin{array}{ll}
w^2(3-2w)
\left\{
\theta(k-w) + \theta(w-k)
\Sigma\left[  \frac{1 - \sqrt{1 - (k/w)^2} }{ 1 + \sqrt{1 - (k/w)^2} }; \alpha(w m_b) \right]
- \theta(w-1) \Sigma\left[w-1;\alpha(m_b)\right]
\right\} &  w < w_M
\\
0        &  w > w_M
\end{array}
\end{equation}
where
\begin{equation}
w_M \, = \, 1 \, + \, \frac{k^2}{4}
\end{equation}
is the maximal hadronic energy above which the spectrum vanishes;
as expected on physical ground, cutting large hadron masses also acts
as an upper cut on hadron energies.
The spectrum is continuous in $w=w_M$ and it
develops large logarithms for $k\rightarrow 0$.
Let us observe that the argument of the first QCD form factor
$\Sigma$ has a similar form to the variable $u$ defined
in eq.~(\ref{defu}). In fact,
\begin{equation}
\left(\frac{k}{w}\right)^2 \,  =\, \left( \frac{\overline{m}_X}{E_X} \right)^2
\end{equation}
is the analogue of the variable $4y$ with $y$ defined in eq.~(\ref{defy}).

\section{Distribution in hadron and electron energies}
\label{sec2energies}

In this section we derive the distribution in the hadron and electron energies
$w$ and $\overline{x}$ by integrating the triple differential distribution (\ref{triplefin})
over $u$. In general, there are two independent energies in the
semileptonic decay (\ref{semilep}).
That is because the hadronic final state $X_u$
is basically a pseudoparticle, i.e. a single entity possessing
an energy $E_X$ and a (variable) mass $m_X$.
We have therefore 3 particles/pseudoparticles in the final state and 3 energies,
related by energy conservation:
\begin{equation}
\label{encons}
x_e \, + \, x_{\nu} + \,w \,= \, 2,
\end{equation}
where
\begin{equation}
x_{\nu} \,=\, \frac{2 E_{\nu}}{m_b}
\end{equation}
and we have written $x_e$ instead of $x$ for aesthetical reasons.
Since the neutrino energy is not usually measured, let us take as independent
energies the electron and the hadron energies.
We have to integrate over $u$ in the range
\begin{equation}
\max[0,w-1] \, \le \, u \, \le \, \min\left[ \frac{ w-\overline{x} }{ \overline{x} },
\frac{ \overline{x} }{ w-\overline{x} } \right].
\end{equation}
As in the previous section, let us use the second method of integration, i.e.
let us omit at first the remainder function.
Since the coefficient function $C(x,w;\,\alpha)$ does not depend on $u$,
the integration only involves the form factor $\sigma$ --- a complementary
situation with respect to the one in the previous section --- and we obtain:
\begin{eqnarray}
\label{inw}
\frac{1}{\Gamma}\frac{d^2\Gamma}{dx dw} & = &
C[\overline{x},w;\,\alpha(m_b)]
\Big\{
\, \theta(2\overline{x}-w)
\Sigma\left[
\frac{ w-\overline{x} }{ \overline{x} } ; \, \alpha(w\,m_b)
\right]
+\theta(w-2\overline{x})
\Sigma\left[
\frac{ \overline{x} }{ w-\overline{x} } ; \, \alpha(w\,m_b)
\right] +
\nonumber\\
&&~~~~~~~~~~~~~~~~~~  - \, \theta(w-1) \Sigma\Big[w-1;\,\alpha(m_b) \Big]
\Big\} \, + \, \cdots,
\end{eqnarray}
where the dots denote non logarithmic terms to be included later.
The decay (\ref{semilep}) involves an hadronic subprocess with a
heavy quark decaying into a light quark evolving later into a jet.
Hadron dynamics is therefore symmetric under the exchange of the
electron and the neutrino momenta, since it is ``blind'' to $W$
decay. That is clearly seen by expressing $w$ through $x_{\nu}$ by
means of eq.~(\ref{encons}):
\begin{eqnarray}
\frac{1}{\Gamma}\frac{d^2\Gamma}{dx_e dx_{\nu}} &=&
C[x_e,x_{\nu};\,\alpha(m_b)]
\Bigg\{ \,
\theta(x_{\nu}-x_e)
\Sigma\left[
\frac{ 1 - x_{\nu} }{ 1-x_e } ; \, \alpha((2-x_e-x_{\nu})m_b)
\right] +
\nonumber\\
& & ~~~~~~~~~~~~~~~~~~~~~
+ \, \theta(x_e-x_{\nu})
\Sigma\left[
\frac{  1-x_e }{ 1 - x_{\nu} } ; \, \alpha((2-x_e-x_{\nu})m_b)
\right] +
\nonumber\\
& & ~~~~~~~~~~~~~~~~~~~~~
- \, \theta(1-x_e-x_{\nu}) \Sigma \Big[1-x_e-x_{\nu};\,\alpha(m_b) \Big]
\Bigg\} \, + \, \cdots.
\end{eqnarray}
Soft-gluon dynamics --- i.e. the expression above in curly brackets ---
is symmetric under exchange of $x_e$ with $x_{\nu}$.
The coefficient function $C[x_e,x_{\nu};\,\alpha(m_b)]$ however is not symmetric
under the exchange of the lepton energies because it does depend
on the whole process, involving the decay of the $W$ boson into the
lepton pair, and not only on the hadronic subprocess.

To proceed with resummation, however, let us go back to the more familiar
variable $w$, i.e. to eq.~(\ref{inw}).
Large logarithms can in principle be obtained by sending to zero
the argument of any of the QCD form factors $\Sigma$'s entering (\ref{inw}),
i.e. in the following three cases:
\begin{equation}
\label{3limits}
          1.~w - \overline{x} \, \rightarrow \, 0;
~~~~~~~~~ 2.~\overline{x} \, \rightarrow \, 0;
~~~~~~~~~ 3.~w \, \rightarrow \, 1^+.
\end{equation}
The coefficient function $C[\overline{x},w;\,\alpha(m_b)]$ vanishes in the first
limit as $O(w-\overline{x})$, implying that in this case there are actually no large
logarithms.
This limit corresponds to $E_{\nu}\rightarrow m_b/2$, a point
where the tree-level spectrum vanishes suppressing soft-gluon effects.
The only relevant limits are therefore the second and the third ones.
It is therefore natural to write a factorization formula
dropping the form factor not associated to large logarithms:
\begin{equation}
\label{mainris}
\frac{1}{\Gamma}\frac{d^2\Gamma}{dx dw} \,=\,
C[\overline{x},w;\,\alpha(m_b)]
\left\{
\Sigma\left[ \overline{x}/w;\,\alpha(w\, m_b) \right]
\, - \,\theta(w-1) \, \Sigma\Big[w-1;\,\alpha(m_b) \Big]
\right\}
\, + \, \cdots.
\end{equation}
We have taken the limit $\overline{x}\rightarrow 0$ in the theta functions
containing $\overline{x}$ in the argument.

Let us consider separately the cases $w\le 1$ and $w>1$.
In the simpler case $w\le 1$
\footnote{
Note that this case is a ``complication'' of the analogous case for the single
distribution in $w$, where the integration over $\overline{x}$ has been made
and therefore there are no large logarithms of $\overline{x}$.}
there is a single form factor and one can write a factorized expression of the form:
\begin{equation}
\frac{1}{\Gamma}\frac{d^2\Gamma}{dx dw} \,=\,
C_L\left(\overline{x},w;\,\alpha\right)
\Sigma\left[ \overline{x} / w;\,\alpha(w\, m_b) \right]
\, + \, d_<(w,\overline{x};\,\alpha)~~~~~~~~~~~~~~~(w<1).
\end{equation}
We require that the remainder function vanishes for
$\overline{x}\rightarrow 0$:
\begin{equation}
\lim_{\overline{x}\rightarrow 0}d_<(w,\overline{x};\,\alpha) \, = \, 0.
\end{equation}
The coefficient function $C_L\left(\overline{x},w;\,\alpha\right)$
can be taken as:
\begin{eqnarray}
C_L^{(0)}(w,\overline{x}) &=& 12 (w-\overline{x})(1+\overline{x}-w);
\\
C_L^{(1)}(w,\overline{x}) &=& \frac{C_F}{\pi} 12 (w-\overline{x})(1+\overline{x}-w)
\left[
{\rm Li}_2(w) + \log w \log(1-w)
-\frac{3}{2}\log w - \frac{w \log w}{2(1-w)} - \frac{35}{8}
\right].
\end{eqnarray}
In $C_L^{(0)}(w,\overline{x})$ we have put the factor
$12(w-\overline{x})(1+\overline{x}-w)$, equal to the spectrum in
lowest order, in order to have a vanishing remainder function in
$O(\alpha^0)$: this is a non minimal choice, since the minimal
choice would imply to set $\overline{x}=0$ in the coefficient
function. We have inserted a similar factor also in
$C_L^{(1)}(w,\overline{x})$, in order to have a simple
multiplicative form of the correction \footnote{We could have
taken as coefficient function the original one
$C(\overline{x},w;\,\alpha)$ as well, which however does not
always contain the factor $12
(w-\overline{x})(1+\overline{x}-w)$.}. As in previous cases, by
matching with the full $O(\alpha)$ result \cite{ndf}, we determine
the remainder function
\begin{equation}
d_<(w,\overline{x};\,\alpha) \,=
\, \alpha\, d_<^{(1)}(w,\overline{x})
+ \, \alpha^2\, d_<^{(2)}(w,\overline{x})
\, + \, O(\alpha^3).
\end{equation}
Omitting the over-all factor $C_F/\pi$,
we obtain for the leading contribution:
\begin{eqnarray}
d_<^{(1)}(w,\overline{x})
&=&-\frac{1}{10}\left( w - \overline{x} \right) \,
       \overline{x}\,
       \big( -210 + 280\,w - 10\,w^2 + 2\,w^3 -
         60\,\overline{x} -
         125\,w\,\overline{x} -
         7\,w^2\,\overline{x} +
         15\,{\overline{x}}^2 \,+\nonumber\\
&+&         32\,w\,{\overline{x}}^2 -
         15\,{\overline{x}}^3 \big)\,+\nonumber\\
&+& \frac{1}{5\,\left( -1 + w \right) }\big( -45\,w + 60\,w^2 -
       20\,w^3 + 10\,w^4 - 6\,w^5 + w^6 -
       15\,\overline{x} +
       135\,w\,\overline{x} -
       255\,w^2\,\overline{x} +\,\nonumber\\
&+&       85\,w^3\,\overline{x} +
       25\,w^4\,\overline{x} -
       5\,w^5\,\overline{x} -
       15\,{\overline{x}}^2 +
       45\,w\,{\overline{x}}^2 +
       75\,w^2\,{\overline{x}}^2 -
       85\,w^3\,{\overline{x}}^2 +
       10\,w^4\,{\overline{x}}^2 \big) \,
     \log w\,+\nonumber\\
&-&
  6\,\left( -1 + w - \overline{x} \right) \,
   \left( w - \overline{x} \right) \,
   \log^2 w
\,+\,
  6\,\left( -1 + w - \overline{x} \right) \,
   \left( w - \overline{x} \right) \,
   \log^2 (w - \overline{x}) \,+\nonumber\\
&-& \frac{1}{5}\left( w -
       \overline{x} \right) \,
     \big( 45 - 15\,w + 5\,w^2 - 5\,w^3 + w^4 +
       15\,\overline{x} -
       10\,w\,\overline{x} +
       15\,w^2\,\overline{x} -
       4\,w^3\,\overline{x} +
       5\,{\overline{x}}^2 \,+\nonumber\\
&-&
       15\,w\,{\overline{x}}^2 +
       6\,w^2\,{\overline{x}}^2 +
       5\,{\overline{x}}^3 -
       4\,w\,{\overline{x}}^3 +
       {\overline{x}}^4 \big) \,
     \log (w - \overline{x}) \,+\nonumber\\
&-&
  \frac{1}{5}\,\overline{x}\,
     \big( 60 - 180\,w + 120\,w^2 +
       60\,\overline{x} -
       15\,w\,\overline{x} -
       45\,w^2\,\overline{x} +
       5\,{\overline{x}}^2 -
       20\,w\,{\overline{x}}^2 +
       10\,w^2\,{\overline{x}}^2 \,+\nonumber\\
&+&
       5\,{\overline{x}}^3 -
       5\,w\,{\overline{x}}^3 +
       {\overline{x}}^4 \big) \,
     \log \overline{x} \,+ \nonumber\\
&+&
  12\,\left( -1 + w - \overline{x} \right) \,
   \left( w - \overline{x} \right) \,\log w\,
   \log \overline{x} -
  12\,\left( -1 + w - \overline{x} \right) \,
   \left( w - \overline{x} \right) \,
   \log (w - \overline{x})\,
   \log \overline{x}~.
\end{eqnarray}
Let us now consider the case $w>1$:
\begin{equation}
\frac{1}{\Gamma}\frac{d^2\Gamma}{dx dw} \,=\,
C\left(\overline{x},w;\,\alpha\right)
\left\{
\Sigma\left[ \overline{x} / w ;\,\alpha(w\, m_b) \right]
\, - \, \Sigma\Big[\Delta w;\,\alpha(m_b) \Big]
\right\}
\, + \, \cdots~~~~~~~~~~(w>1),
\end{equation}
where we have defined
\begin{equation}
\Delta w \, = \, w \, - \, 1 \, > \, 0.
\end{equation}
There are two form factors and
large logarithms can be obtained in the following three kinematical
configurations:
\begin{enumerate}
\item
$\overline{x} \ll \Delta w \sim 1$:
large logarithms of the form $\alpha^n\log^k\overline{x}$ have to be resummed;
\item
$\Delta w \sim \overline{x} \ll 1$:
large logarithms of the form
$\log \Delta w \sim \log\overline{x}$ have to be resummed;
\item
$\Delta w \ll \overline{x} \sim 1$:
large logarithms of the form $\log\Delta w$ have to be resummed.
\end{enumerate}
The first case is kinematically forbidden because
\begin{equation}
\Delta w \, \le \, \overline{x}.
\end{equation}
The second case does not give large logarithms because
the coefficient function $C(\overline{x},w;\,\alpha)$ vanishes linearly in this
limit:
\begin{equation}
C(\lambda\,\overline{x},1+\lambda\,\Delta w;\,\alpha)\, = \, O(\lambda)
~~~~~{\rm for}~\lambda\rightarrow 0.
\end{equation}
The only relevant limit is therefore the third one,
implying that one can drop the form factor
$\Sigma(\overline{x}/w;\,\alpha)$.
We propose then a resummed form for this distribution which is
a generalization of that one for the hadron energy spectrum:
\begin{equation}
\frac{1}{\Gamma}\frac{d^2\Gamma}{dx dw} \,=\,
C_{XW1}\left(\overline{x};\,\alpha\right)
\left\{1 \, - \,
C_{XW2}\left(\overline{x};\,\alpha\right)
\Sigma\Big[\Delta w;\,\alpha(m_b) \Big]\right\}
\, + \, d_>(\Delta w,\overline{x};\,\alpha)
~~~~~~~~~~~~~~~~~(w>1).
\end{equation}
We require that the remainder function vanishes for
$\Delta w\rightarrow 0^+$:
\begin{equation}
\lim_{\Delta w \rightarrow 0^+} d_>(\Delta w,\overline{x};\,\alpha) \, = \, 0.
\end{equation}
The first coefficient function is obtained by imposing the
continuity of the spectrum for $w\rightarrow 1$ from both sides
$w<1$ and $w>1$ and for any $\overline{x}$ \footnote{ This
continuity condition, which involves a single point $w=1$ for the
hadron energy spectrum, involves in this more complicated case the
line $(w=1,\,\bar{x})$.}. We obtain:
\begin{equation}
C_{XW1}(\overline{x};\,\alpha)
 \, = \, C_L\left(\overline{x},1;\,\alpha\right)
\Sigma\left(\overline{x};\,\alpha \right)
\, + \, d_<(1,\overline{x};\,\alpha).
\end{equation}
We can expand in the above equation $\Sigma\left(\overline{x};\,\alpha \right)$
in powers of $\alpha$ (up to first order) because the coefficient function
for $w=1$, $C_L\left(\overline{x},1;\,\alpha\right)$, vanishes linearly for
$\overline{x}\rightarrow 0$, killing the large logarithms in the form factor.
We then obtain:
\begin{equation}
C_{XW1}(\overline{x};\,\alpha)
\,=\, C_{XW1}^{(0)}(\overline{x})
\,+\,\alpha\, C_{XW1}^{(1)}(\overline{x})
\,+\,\alpha^2\, C_{XW2}^{(2)}(\overline{x})
\,+\,O(\alpha^3)
\end{equation}
where
\begin{eqnarray}
C_{XW1}^{(0)}(\overline{x})&=& 12\,(1-\overline{x})\,\overline{x};
\\
C_{XW1}^{(1)}(\overline{x})&=&\frac{C_F}{\pi}\bigg\{
\frac{1}{10}\left( 1 - \overline{x} \right) \,
     \overline{x}\,
     \left( -587 + 192\,\overline{x} -
       47\,{\overline{x}}^2 +
       15\,{\overline{x}}^3 \right)
- \frac{1}{5}\,\overline{x}\,
     \left( 105 - 105\,\overline{x} -
       5\,{\overline{x}}^2 + {\overline{x}}^4
       \right) \,\log \overline{x}\,+\nonumber\\
& -&
  \frac{1}{5}\left( 1 - \overline{x} \right) \,
     \left( 31 + 16\,\overline{x} -
       4\,{\overline{x}}^2 +
       {\overline{x}}^3 + {\overline{x}}^4
       \right) \,\log (1 - \overline{x}) -
  6\,\left( 1 - \overline{x} \right) \,
   \overline{x}\,\log^2 (1 - \overline{x})\,+\nonumber\\
& +&
  12\,\left( 1 - \overline{x} \right) \,
   \overline{x}\,\log (1 - \overline{x})\,
   \log \overline{x} -
  6\,\left( 1 - \overline{x} \right) \,
   \overline{x}\,\log^2 \overline{x} +
  12\,\left( 1 - \overline{x} \right) \,
   \overline{x}\,z(2)
\bigg\} .
\end{eqnarray}
The second coefficient function $C_{XW2}(\overline{x};\,\alpha)$
is obtained by matching with the fixed-order distribution in the limit
$\Delta w\rightarrow 0^+$:
\begin{equation}
C_{XW2}(\overline{x};\,\alpha)
\,=\, C_{XW2}^{(0)}(\overline{x})
\,+\,\alpha\, C_{XW2}^{(1)}(\overline{x})
\,+\,\alpha^2\, C_{XW2}^{(2)}(\overline{x})
\,+\,O(\alpha^3)
\end{equation}
with
\begin{eqnarray}
C_{XW2}^{(0)}(\overline{x})&=& 1;
\\
C_{XW2}^{(1)}(\overline{x})&=&
\frac{C_F}{\pi}\bigg\{
\frac{1}{120} \left( 62 - 192\,\overline{x} +
     47\,{\overline{x}}^2 -
     15\,{\overline{x}}^3 \right) \,+\,
  \frac{1}{60\,\overline{x}}
       \left( 31 + 16\,\overline{x} -
       4\,{\overline{x}}^2 +
       {\overline{x}}^3 + {\overline{x}}^4
       \right) \,\log (1 - \overline{x})\, +\nonumber\\
&+&
  \frac{1}{2}\,\log^2 (1 - \overline{x})\, + \,
  \frac{1}{60\,\left( 1 - \overline{x} \right) }\,
    \left( 105 - 105\,\overline{x} -
       5\,{\overline{x}}^2 + {\overline{x}}^4
       \right) \,\log \overline{x} \, -\,
  \log (1 - \overline{x})\,
   \log \overline{x}\,+\nonumber\\
& + &
  \frac{1}{2}\log^2 \overline{x}
 \bigg\}.
\end{eqnarray}
The remainder function $d_>(w,\overline{x};\,\alpha)$
is obtained by matching with the fixed-order distribution for $\Delta w\sim O(1)$:
\begin{equation}
d_>(w,\overline{x};\,\alpha)
\,=\, \alpha\, d_>^{(1)}(w,\overline{x})
\, + \, \alpha^2\, d_>^{(2)}(w,\overline{x})
\,+\,O(\alpha^3),
\end{equation}
where, omitting the overall factor $C_F/\pi$:
\begin{eqnarray}
d_>^{(1)}(w,\overline{x}) &=&
\frac{1}{10}\left( -1 + w \right) \,
     \left( 1 - \overline{x} \right) \,
     \big( -75 + 142\,w - 7\,w^2 + 2\,w^3 -
       212\,\overline{x} -
       105\,w\,\overline{x} -
       9\,w^2\,\overline{x}\,+\nonumber\\
& +&
       132\,{\overline{x}}^2 +
       39\,w\,{\overline{x}}^2 -
       47\,{\overline{x}}^3 \big)\,+\,
  \frac{1}{5}\left( -1 + w \right) \,
     \big( -4 - 99\,w + w^2 - 4\,w^3 + w^4 +
       140\,\overline{x}\,+\nonumber\\
&+&
       120\,w\,\overline{x} +
       15\,w^2\,\overline{x} -
       5\,w^3\,\overline{x} -
       65\,{\overline{x}}^2 -
       65\,w\,{\overline{x}}^2 +
       10\,w^2\,{\overline{x}}^2 \big) \,
     \log (-1 + w)\,+\nonumber\\
& -&
  6\,\left( -1 + w \right) \,
   \left( w - 2\,\overline{x} \right) \,
   \log^2 (-1 + w) +
  \frac{1}{5}\left( 1 - \overline{x} \right) \,
     \left( 31 + 16\,\overline{x} -
       4\,{\overline{x}}^2 +
       {\overline{x}}^3 + {\overline{x}}^4
       \right) \,\log (1 - \overline{x})\,+\nonumber\\
&+&
  6\,\left( 1 - \overline{x} \right) \,
   \overline{x}\, \log^2 (1 - \overline{x}) +
   6\,\left( -1 + w - \overline{x} \right) \,
   \left( w - \overline{x} \right) \,
   \log^2 (w - \overline{x})\,+\nonumber\\
& -&
  \left( -1 + w \right) \,
   \left( -21\,w + 30\,\overline{x} +
     24\,w\,\overline{x} -
     12\,{\overline{x}}^2 -
     9\,w\,{\overline{x}}^2 -
     2\,{\overline{x}}^3 +
     2\,w\,{\overline{x}}^3 - {\overline{x}}^4
     \right) \,\log \overline{x}\,+\nonumber\\
& -&
  12\,\left( 1 - \overline{x} \right) \,
   \overline{x}\,\log (1 - \overline{x})\,
   \log \overline{x} +
  6\,\left( -1 + w \right) \,
   \left( w - 2\,\overline{x} \right) \,
   \log^2 \overline{x}\,+\nonumber\\
& -&
   \frac{1}{5}\left( w -
            \overline{x} \right) \,
          \big( 45 - 15\,w + 5\,w^2 - 5\,w^3 +
            w^4 + 15\,\overline{x} -
            10\,w\,\overline{x} +
            15\,w^2\,\overline{x} -
            4\,w^3\,\overline{x} +
            5\,{\overline{x}}^2\,+\nonumber\\
& -&
            15\, w\,{\overline{x}}^2 +
            6\, w^2\,{\overline{x}}^2 +
            5\,{\overline{x}}^3 -
            4\,w\,{\overline{x}}^3 +
            {\overline{x}}^4 \big) \,\log (w - \overline{x})\,+\nonumber\\
 &-& 12\,\left( -1 + w -
        \overline{x} \right)
      \left( w - \overline{x} \right)\,
      \log \overline{x}\,  \log (w - \overline{x})\,.
\end{eqnarray}
To summarize, we have presented a complete NLO resummation
of the distribution in the hadron and electron energies
$w$ and $x$, which is a generalization of the resummation
of the hadron energy spectrum of the previous section.
Resummation takes a different form in the cases $w\le 1$
and $w>1$.
In the first case there is a series of threshold logarithms
of the form
\begin{equation}
\alpha^n \, \log^k \frac{\overline{x}}{w}
~~~~~~~~~~~~~~~~~~~~~~~~~~~~~~~~~~~~~~~~~~~~~~~~~~~~~~(w<1),
\end{equation}
while in the second case the infrared logarithms are of the form
\begin{equation}
\alpha^n \, \log^k (w-1)
~~~~~~~~~~~~~~~~~~~~~~~~~~~~~~~~~~~~~~~~~~~~~~~~~~~~(w>1).
\end{equation}
Unlike the distribution in sec.~\ref{hadrvars}, we have here a
differential distribution involving the partially-integrated form
factor $\Sigma$.

\section{Conclusions}
\label{concl}

It is a rather old idea that semi-inclusive $B$ decays can be related to
each other because of some universal long-distance component \cite{altetal}.
We have presented in this paper a critical analysis of this idea, based on
a resummation formula for the triple differential distribution
in the semileptonic decay (\ref{semilep}).
Long-distance effects manifest themselves  in perturbation theory in the form of
series of large infrared logarithms, coming from the multiple
emission of soft and/or collinear gluons.
The universality of long-distance effects has
therefore to show up in perturbation theory in the form of
identical series of large logarithms in different distributions.
Semi-inclusive $B$ decays have been defined in all generality as
decays of the form
\begin{equation}
\label{allafine}
B \, \rightarrow \, X_q \,+\, {\rm (non~QCD~partons)},
\end{equation}
in the kinematical region close to the threshold $m_X=0$, i.e. for
\begin{equation}
m_X \, \ll \, E_X.
\end{equation}
We have shown that semileptonic distributions are naturally divided  into two classes.

The first class contains distributions which are not integrated over the hadronic
energy $E_X$ and consequently have a long-distance structure similar to the one in
radiative decays (\ref{bsgamma}).
These are the (simpler) distributions to attack and
have been treated in this paper. We have resummed to next-to-leading order:
\begin{enumerate}
\item the distribution in the hadronic energy $E_X$ and in the
variable $u$ defined in sec.~(\ref{secthresum}), which is
basically the ratio $m_X^2/(4E_X^2)$, i.e. the hadron invariant
mass squared in unit of the hard scale; \item the hadron energy
distribution, which is a case of the so-called Sudakov shoulder.
This is the only single distribution which can be related to the
radiative decay via short-distance factors only. The large
logarithms which appear in this distribution are indeed equal to
the ones which appear in the radiative decay (\ref{bsgamma}). We
have studied in detail the relation between the hadron energy
spectrum and the photon spectrum in the radiative decay. It is
remarkable that the large logarithms in the hadron energy spectrum
occur at $E_X=m_b/2$, i.e. when the hard scale $Q=2E_X$ equals
$m_b$, as in the radiative decays; \item the distribution in the
hadron and in the charged lepton energies, which contains two
different classes of large logarithms according to the cases $w\le
1$ or $w>1$. The resummation of this distribution is the most
complicated  and is a generalization of the resummation of the
hadron energy spectrum.
\end{enumerate}
The second class contains semileptonic distributions in which the
hadronic energy is integrated over, such as for example the
hadronic mass distribution or the charged lepton energy
distribution. These distributions have a complicated logarithmic
structure, which is not simply related to the one in the radiative
decay and there is not a pure short-distance relation with the
radiative decay spectrum. The resummation of these distributions
to NLO is presented in \cite{wip}.

\vskip 0.5truecm

\centerline{\bf Acknowledgments}

One of us (U.A.) wishes to thank R. Faccini for discussions.


\begin{thebibliography}{99}

\bibitem{cattren1}
S. Catani and L. Trentadue, Nucl. Phys. B 327, 323 (1989).

\bibitem{dok}
Y. Dokshitzer et al., {\it Basics of Perturbative QCD}, Editions
Frontieres, Paris (1991); R. Ellis, W. Stirling and B. Webber,
{\it QCD and Collider
Physics}, Cambridge University Press, Cambridge (1996).

\bibitem{me}
U. Aglietti, Nucl. Phys. B 610, 293 (2001), (hep-ph/0104020 v3).

\bibitem{thrust}
S. Catani, L. Trentadue, G. Turnock and B. Webber, Nucl. Phys. B 407, 3 (1993).

\bibitem{wip}
U. Aglietti, G. Ferrera and G. Ricciardi, hep-ph/0509095v1.

\bibitem{sudshould}
S. Catani and B. Webber, J. High Energy Phys. 10, 005 (1997) (hep-ph/9710333).

\bibitem{enspec}
A. Czarnecki, M. Jezabek and J. Kuhn, Acta Phys. Polon. B 20, 961 (1989).

\bibitem{ndf}
F. De Fazio and M. Neubert, J. High Energy Phys. 06, 017 (1999) (hep-ph/9905351).

\bibitem{uanalog}
W. Bernreuther et al., Nucl. Phys. B 706, 245 (2005); B 712, 229 (2005).

\bibitem{ucg}
U. Aglietti, M. Ciuchini and P. Gambino, Nucl. Phys. B 637 427-444 (2002) (hep-ph/0204140).

\bibitem{akhouri}
R. Akhoury and I. Rothstein, Phys. Rev. D 54, 2349 (1996) (hep-ph/9512303).

\bibitem{dileptonspec}
A. Czarnecki and K. Melnikov, Phys. Rev. Lett. 88, 131801 (2002)
(hep-ph/0112264).

\bibitem{backtox}
S. Catani, M. Mangano, P. Nason and L. Trentadue, Nucl. Phys. B 478, 273 (1996)
(hep-ph/9604351).

\bibitem{catani}
For an introduction see for example: S. Catani, Proceedings of QCD
{\it Euroconference} 96, Montpellier, France, July 1996
(hep-ph/9709503).

\bibitem{sterman}
G. Sterman, Nucl. Phys. B 281, 310 (1987).

\bibitem{cattren2}
S. Catani and L. Trentadue, Nucl. Phys. B 353, 183 (1991).

\bibitem{model}
For a recent discussion on this problem see for example:
U. Aglietti and G. Ricciardi, Phys. Rev. D 70, 114008 (2004) (hep-ph/0204125v1).

\bibitem{catweb}
S. Catani and B. Webber and G. Marchesini, Nucl. Phys. B 349, 635 (1991).

\bibitem{beta2}
O. Tarasov, A. Vladimirov and A. Zharkov, Phys.~Lett.~B 93, 429 (1980);
S. Larin and J. Vermaseren, Phys.~Lett.~B 303, 334 (1993) (hep-ph/9302208).

\bibitem{beta3}
T. Van Ritbergen, J. Vermaseren and S. Larin, Phys.~Lett.~B 400,
379 (1997) (hep-ph/9701390); K. Chetyrkin, B. Kniehl and M.
Steinhauser, Phys. Rev. Lett. 79, 2184 (1997) (hep-ph/9706430);
M.~Czakon,  Nucl. Phys. B 710, 485 (2005) (hep-ph/0411261).

\bibitem{proclecce}
U. Aglietti, Proceedings of the XV Italian Meeting on High Energy Physics (IFAE),
Lecce, Italy, 23-26 April 2003.

\bibitem{conlucastud}
U. Aglietti, R. Sghedoni and L. Trentadue,
Phys. Lett. B 522,  83 (2001) (hep-ph/0105322);
Phys. Lett. B 585, 131 (2004) (hep-ph/0310360).

\bibitem{demilio}
S. Catani, E. D'Emilio and L. Trentadue, Phys. Lett. B 211, 335 (1988).

\bibitem{lucavecchio}
J. Kodaira and L. Trentadue, Phys. Lett. B 123, 335 (1983).

\bibitem{altetal}
G. Altarelli, N. Cabibbo, G. Corb\`o, L. Maiani and G. Martinelli,
Nucl. Phys. B 208, 365 (1982).

\bibitem{A2}
J. Kodaira and L. Trentadue, SLAC-PUB-2934 (1982);
Phys. Lett. B 112, 66 (1982).

\bibitem{twoloopkernels}
G. Curci, W. Furmansky and R. Petronzio, Nucl. Phys. B 175, 27 (1980).

\bibitem{A3}
S. Moch, J. Vermaseren and A. Vogt, Nucl. Phys. B 688, 101 (2004)
(hep-ph/0402192);
Nucl. Phys. B 691, 129 (2004) (hep-ph/0404111v1).

\bibitem{korrad}
G. Korchemsky and A. Radyushkin, Nucl. Phys. B 283, 342 (1987).

\bibitem{B2}
S. Moch, J. Vermaseren and A. Vogt, Nucl. Phys. B 646, 181 (2002) (hep-ph/0209100v1).

\bibitem{checknnlo}
S. Moch, J. Vermaseren and A. Vogt, hep-ph/0504242.

\bibitem{D2first}
G. Korchemsky and G. Marchesini, Nucl. Phys. B 406, 225 (1993).

\bibitem{neub}
M. Neubert, Eur. Phys. J. C 40, 165 (2005) (hep-ph/0408179).

\bibitem{fragmentation}
K. Melnikov and A. Mitov, Phys. Rev. D 70, 034027 (2004)
(hep-ph/0404143v2).

\bibitem{gardi0}
E. Gardi, JHEP 0502, 53 (2005) (hep-ph/0501257).

\bibitem{jafferandall}
R. Jaffe and L. Randall, Nucl. Phys. B 412, 79 (1994),
(hep-ph/9306201).

\bibitem{O7twoloops}
K. Melnikov and A. Mitov, Phys. Lett. B 620, 69 (2005)
(hep-ph/0505097v1) .

\bibitem{gardi}
J. Anserson and E. Gardi, JHEP 0506,30 (2005)
(hep-ph/0502159v2).

\bibitem{ugnnlo}
U. Aglietti and G. Ricciardi, Phys. Rev. D 66, 074003 (2002) (hep-ph/0204125 v1).

\bibitem{vanervogt}
W. Van Neerven and A. Vogt, Phys. Lett. B 490, 111 (2000) (hep-ph/0007362).

\end{thebibliography}
\end{document}